\documentclass[acmsmall]{acmart}

\usepackage{xcolor}
\usepackage{colortbl}
\usepackage[linesnumbered,ruled,vlined]{algorithm2e}
\usepackage{amsmath}
\usepackage{hyperref}

\DeclareMathOperator*{\argmax}{arg\,max}

\newcommand{\titleicon}{%
  \ifnum\value{page}=1
    \includegraphics[width=1.5em]{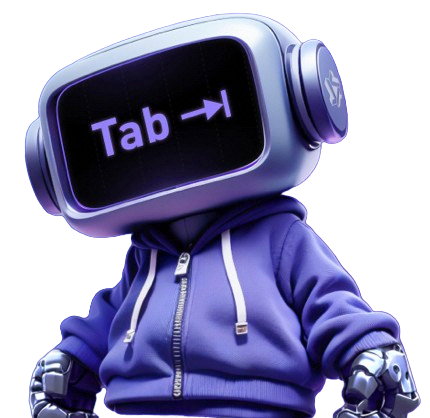}%
  \fi
}

\newcommand{\greencomment}[1]{\textcolor{green!50!black}{\tcp*{#1}}}

\AtBeginDocument{%
  }

\setcopyright{none}
\makeatletter
\renewcommand\@formatdoi[1]{\ignorespaces}
\makeatother
\renewcommand\footnotetextcopyrightpermission[1]{} 
\begin{document}

\title{Lingma SWE-GPT~\titleicon:\\ An Open Development-Process-Centric Language Model for Automated Software Improvement}

\author{Yingwei Ma}
\author{Rongyu Cao}
\author{Yongchang Cao}
\author{Yue Zhang}
\author{Jue Chen}
\author{Yibo Liu}
\author{Yuchen Liu}
\author{Binhua Li}
\author{Fei Huang}
\author{Yongbin Li}
\authornote{Corresponding Author.}
\email{mayingwei.myw@alibaba-inc.com}
\affiliation{%
  \institution{Tongyi Lab, Alibaba Group}
  \city{Beijing}
  \country{China}
}

\renewcommand{\shortauthors}{Yingwei Ma, Rongyu Cao YongChang Cao et al.}

\begin{abstract}
Large language models (LLMs) have demonstrated remarkable performance in code generation, significantly enhancing the coding efficiency of developers. Recent advancements in LLM-based agents have led to significant progress in end-to-end automatic software engineering (ASE), particularly in software maintenance (e.g., fixing software issues) and evolution (e.g., adding new features). Despite these encouraging advances, current research faces two major challenges. First, state-of-the-art performance primarily depends on closed-source models like GPT-4, which significantly limits the technology's accessibility, and potential for customization in diverse software engineering tasks. This dependence also raises concerns about data privacy, particularly when handling sensitive codebases. Second, these models are predominantly trained on static code data, lacking a deep understanding of the dynamic interactions, iterative problem-solving processes, and evolutionary characteristics inherent in software development. Consequently, they may face challenges in navigating complex project structures and generating contextually relevant solutions, which can affect their practical utility in real-world scenarios.

To address these challenges, our study adopts a software engineering perspective. We recognize that real-world software maintenance and evolution processes encompass not only static code data but also developers' thought processes, utilization of external tools, and the interaction between different functional personnel. Our objective is to develop an open-source large language model specifically optimized for software improvement, aiming to match the performance of closed-source alternatives while offering greater accessibility and customization potential. Consequently, we introduce the \textbf{Lingma SWE-GPT} series, comprising Lingma SWE-GPT 7B and Lingma SWE-GPT 72B. By learning from and simulating real-world code submission activities, Lingma SWE-GPT systematically incorporates the dynamic interactions and iterative problem-solving inherent in software development process—such as repository understanding, fault localization, and patch generation—thereby achieving a more comprehensive understanding of software improvement processes. We conducted experimental evaluations using SWE-bench-Verified benchmark (comprising 500 real GitHub issues), recently proposed by OpenAI. The results demonstrate that \textbf{Lingma SWE-GPT 72B successfully resolves 30.20\% of the GitHub issues}, marking a significant improvement in automatic issue resolution (22.76\% relative improvement compared to Llama 3.1 405B), approaching the performance of closed-source models (31.80\% issues of GPT-4o resolved). Notably, Lingma SWE-GPT 7B resolves 18.20\% of the issues, surpassing the 17.20\% resolution rate of Llama 3.1 70B, highlighting the potential for applying smaller models to ASE tasks.
\end{abstract}






\maketitle

\section{Introduction}

Automated software engineering (ASE) has long been a vision pursued by both the software engineering (SE) and artificial intelligence (AI) communities. Recent advancements in large language models (LLMs) have shown significant potential in advancing this field. Initially, the HumanEval benchmark~\cite{humaneval} was developed to assess LLMs' capabilities in function-level code generation. Both closed-source (e.g., GPT-4o~\cite{gpt4o}) and open-source models (e.g., Llama 3.1 405B~\cite{llama3.1}) have performed well on these tasks, solving more than 90\% of problems. However, function-level code generation represents only a fraction of the challenges encountered in real-world software development. 

To evaluate LLMs' capabilities in more realistic scenarios, the SWE-bench benchmark~\cite{jimenez2023swe} series was introduced. The evaluation process in SWE-bench is designed to simulate real-world software improvement tasks: given a natural language description of an issue and the corresponding Github repository, the model is expected to generate a patch that resolves the issue. This approach tests not only code generation capabilities but also the model's ability to understand real-world software issue, locate relevant code across multiple files, and make appropriate modifications while maintaining the integrity of the entire codebase. In response to these challenges, both SE and AI  communities have focused on developing sophisticated software engineering agents~\cite{chen2024coder, liu2024marscode, cognitionai2023devin, ma2024understand}. Systems such as SWE-agent~\cite{yang2024sweagent} and AutoCodeRover~\cite{autocoderover} exemplify these approaches, leveraging the general capabilities of LLMs to iteratively and autonomously formulate plans, execute tool calls, and observe feedback. For example, SWE-agent employs an agent-computer interface to execute operations like opening files, searching code lines, and running tests. In contrast, approaches like Agentless~\cite{xia2024agentless} utilize LLMs within predefined workflows that guide the problem-solving process through fixed, expert-designed steps. While this method limits the autonomy and generality of LLM, it also shows promising results by effectively structuring tasks. However, our analysis reveals that current ASE approaches still face significant limitations due to two primary factors. 

\textbf{Over-reliance on Closed-Source Models.} The top-performing submissions on the SWE-bench~\cite{jimenez2023swe} leaderboard are exclusively based on closed-source models like GPT-4/4o~\cite{achiam2023gpt4, gpt4o} and Claude 3.5 Sonnet~\cite{claude3.5}. In contrast, submissions utilizing open-source models, such as SWE-Llama 13B, solve only 1.00\% of the problems. Although recent research has shown progress in open-source model performance—with Qwen2 72B instrut achieving 9.34\% on SWE-bench Lite~\cite{liu2024codexgraph}—this advancement, while encouraging, still lags significantly behind closed-source models. Moreover, considering the privacy of user code repositories in software development, utilizing closed-source models for code analysis and modification may raise security concerns. This factor restricts the applicability of closed-source models in real-world software engineering scenarios, highlighting the urgent need for high-performance open-source models.

\textbf{Lack of Comprehensive Development Process Data.} General-purpose and code-specific large models are typically trained on vast amounts of static code data~\cite{hui2024qwen2.5-coder, zhu2024deepseekcoder, llama3.1}. While this approach has significantly enhanced their code generation capabilities, it overlooks the dynamic interactions and development process crucial to comprehensively understanding real-world software engineering practices. Real-world software improvement encompasses a complex reasoning cycle of issue identification, tool utlization, code navigation, and iterative problem-solving, which is not captured by merely training on static codebases. Although some existing models~\cite{muennighoff2023octopack, lozhkov2024starcoder} utilize pull request submission processes recorded on GitHub, these data only include commit messages and final patches, lacking the detailed thought processes and reasoning of developers during problem-solving. Furthermore, current training methods primarily focus on isolated code snippets or single files, neglecting broader project structures and cross-file information. This results in models lacking a global perspective when dealing with complex software systems, making it challenging to accurately understand and handle cross-module interactions.

\begin{figure*}
    \centering
    \includegraphics[scale=0.41]{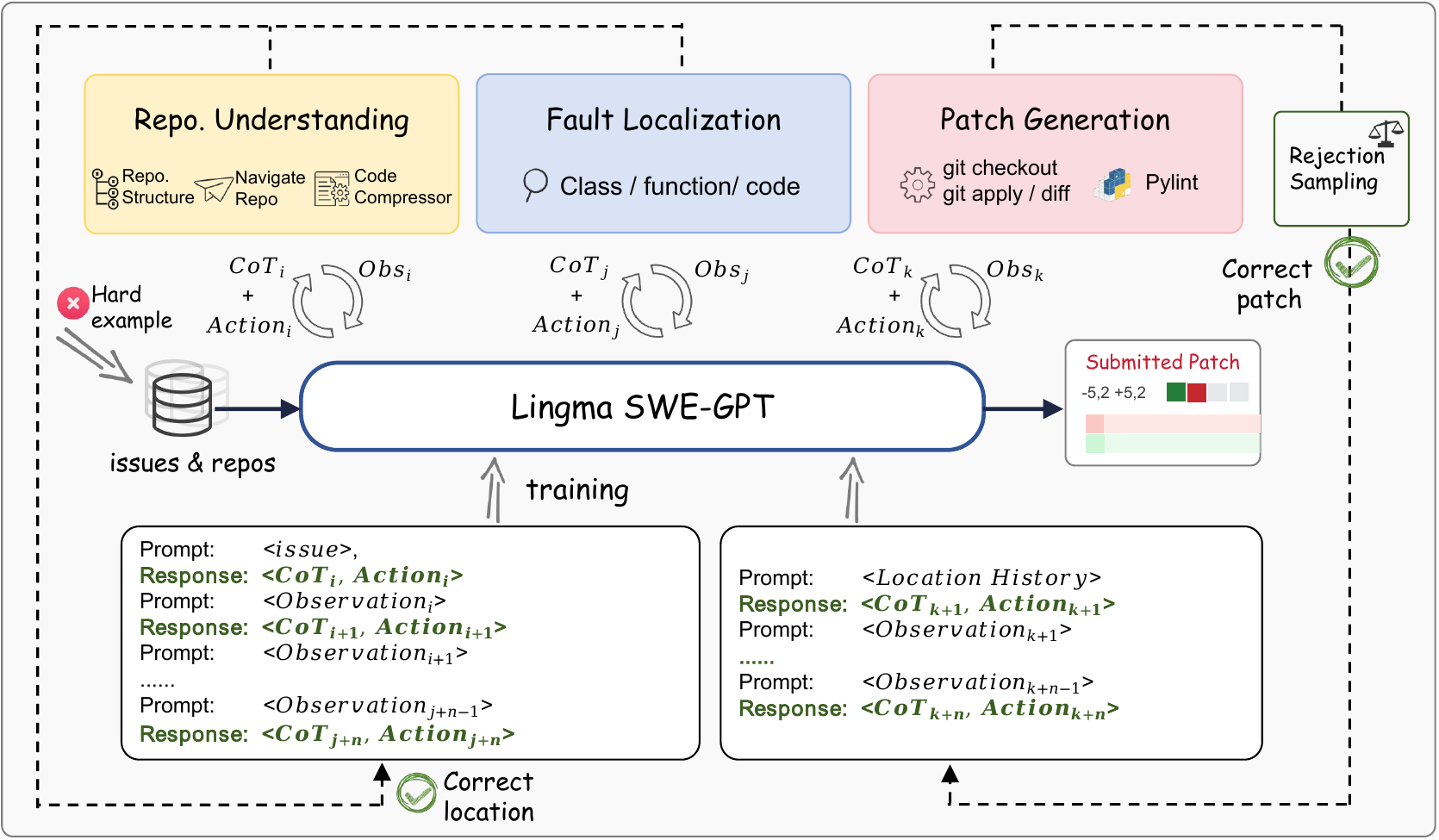}
    \caption{Overview of an iterative development-process training framework for automated software improvement.}
    \label{fig:training}
\end{figure*}

\textbf{Our Approach.} To address these challenges, we introduce the Lingma SWE-GPT series, which includes models of two size, 7B and 72B, specifically designed for software improvement. As shown in Figure \ref{fig:training}, our approach simulates the software improvement process through a three-stage process: repository understanding, fault localization, and patch generation. Each stage operates in a Chain-of-Thought (CoT) manner. To address real-world GitHub issues, Lingma SWE-GPT initially conducts a comprehensive analysis of the software repository, examining the codebase hierarchically from the overall file structure to specific classes and functions. Specifically, Lingma SWE-GPT identifies a set of potentially relevant files based on the natural language description of the issue and the repository's directory tree structure. It then identifies relevant classes and functions based on the skeletons of these files and formulates a plan for issue resolution.
Following this repository understanding, Lingma SWE-GPT retrieves pertinent code context by invoking specialized search APIs (e.g., search\_func(\textit{'resize'})). These APIs, leveraging Abstract Syntax Tree analysis, extract contextual information such as method and class implementations from the codebase. The model iteratively refines its understanding of both the issue and the repository, strategically selecting which APIs to use in subsequent iterations until potential fault locations are identified. 
In the concluding phase, Lingma SWE-GPT generates and applies patches to address the localized issues. This phase encompasses concrete solution generation, code replacement, and iterative debugging processes based on syntax validation and git operations, ensuring the development of applicable patches.
To further enhance the model's capabilities, we employ a development process-centric iterative training strategy. This involves optimizing model performance through maximizing the conditional probability of generating development process outputs, including thought processes, tool utilization record and final results. By incorporating curriculum learning, the model progressively tackles increasingly complex tasks, establishing a robust foundation on simpler ones. Additionally, we implement a rejection sampling process to ensure the quality of synthesized data, selectively retaining high-quality instances that closely mimic real-world software development practices.

Our extensive evaluation on SWE-bench Verified and SWE-bench Lite demonstrates the effectiveness of Lingma SWE-GPT: the 72B version successfully resolves 30.20\% of issues on SWE-bench Verified, marking a significant improvement over existing open-source models(22.76\% relative improvement compared to Llama 3.1 405B) and approaching the performance of leading closed-source alternatives (31.80\% issues of GPT-4o resolved). Additionally, the 7B version achieves an impressive 18.20\% success rate, surpassing the 17.20\% resolution rate of Llama 3.1 70B, showcasing the potential of smaller, more efficient models in automated software engineering tasks.

\textbf{Contributions.} In summary, we make the following novel contributions:
\begin{itemize}
\item We introduce Lingma SWE-GPT, a novel series of open-source large language models specifically optimized for automated software improvement\footnote{\url{https://github.com/LingmaTongyi/Lingma-SWE-GPT}}.
\item We propose a development process-centric training approach that captures the dynamic nature of software engineering, including tool utilizing, reasoning, and interactive problem-solving capabilities.
\item We demonstrate the effectiveness of our approach through comprehensive evaluations on SWE-bench Lite and Verified, showing significant improvements over existing open-source models and competitive performance against closed-source alternatives.
\item We provide insights into the model's fault localization capabilities and performance consistency, offering valuable directions for future research in AI-assisted software engineering.
\end{itemize}

\section{Related Work}
\subsection{Large Language Models for Code}


Generative models such as ChatGPT have exhibited significant capabilities in code generation and comprehension. These models have substantially impacted various aspects of software engineering, enabling tasks such as code generation~\cite{ma2023training, pan2024codev, jiang2023automatic, zhu2024domaineval, xu2024cruxeval} from natural language requirements, test generation~\cite{liu2024your, xue2024selfpico, xia2024fuzz4all, li2024dllens}, and code editing and refactoring~\cite{li2023codeeditor, chakraborty2021multi, shirafuji2023refactoring, alomar2024refactor, zhang2024lpr, zhang2023lampr}. Furthermore, developers and researchers have applied these models to more complex software engineering tasks, including code translation~\cite{pan2023understanding, liu2024mftcoder,gong2024ast}, debugging~\cite{chen2023teaching, ding2024cycle, shi2024code, ni2024next, yan2024better}, and automated program repair~\cite{jimenez2023swe, autocoderover, li2023two}.


The effectiveness of these models can be attributed to their pretraining on extensive general-purpose data and open-source code repositories. This extensive training has facilitated the development of sophisticated code generation and reasoning capabilities. Notable examples of such models include GPT-4~\cite{achiam2023gpt4}, Claude 3.5 Sonnet \cite{claude3.5}, CodeX~\cite{humaneval}, Code Llama~\cite{codellama}, StarCoder~\cite{lozhkov2024starcoder}, DeepSeek-Coder \cite{zhu2024deepseekcoder}, Qwen2.5 Coder~\cite{hui2024qwen2.5-coder}, and CodeGemma~\cite{team2024codegemma}. These models have shown considerable proficiency in comprehending and generating code across various programming languages and paradigms. In addition to pretrained LLMs, researchers have developed instruction-tuned models specifically tailored for code-related tasks. Examples of such models include CodeLlama-Instruct~\cite{codellama}, Pangu-coder~\cite{shen2023pangu}, WizardCoder~\cite{luo2023wizardcoder}, Magicoder \cite{wei2024magicoder}, WaveCoder~\cite{yu2023wavecoder} and Opencodeinterpreter~\cite{zheng2024opencodeinterpreter}. These models undergo additional fine-tuning with carefully curated instructions, enhancing their performance on specific downstream code-related tasks. Despite their advanced capabilities, current LLMs are primarily trained on static code data, limiting their understanding of the dynamic interactions nature of software development. This paper proposes developing models that can simulate these dynamic aspects, including reasoning about tool usage and mimicking thought processes, to better address the challenges of real-world software engineering tasks.

\subsection{LLM-based Software Engineering Agents}

In recent years, LLM-based AI agents have advanced the development of automatic software engineering (ASE). AI agents improve the capabilities of project-level software engineering tasks through running environment awareness~\citep{hong2023metagpt, wang2024codeact, kong2024contrastrepair, xie2024pet}, planning \& reasoning~\citep{wang2024codeact, cognitionai2023devin, luo2024repoagent}, and tool construction~\citep{zhang2024codeagent, lee2024unified, xue2023acwrecommender, ma2023mulcs, huang2023towards}. Surprisingly, Devin~\citep{cognitionai2023devin} is a milestone that explores an end-to-end LLM-based agent system to handle complex SE tasks. Concretely, it first plans the requirements of users, then adopts the editor, terminal and search engine tools to make independent decisions and reasoning, and finally generates codes to satisfy the needs of users in an end-to-end manner. Its promising designs and performance swiftly ignited unprecedented attention from the SE and AI community to automatic software engineering (ASE)~\citep{yang2024sweagent, autocoderover, xia2024agentless, ma2024understand, liu2024large, liu2024marscode, chen2024coder, liu2024codexgraph, zhu2024moss}. For example, SWE-agent~\citep{yang2024sweagent} carefully designs an Agent Computer Interface (ACI) to empower the SE agents capabilities of creating \& editing code files, navigating repositories, and executing programs. Besides, AutoCodeRover~\citep{autocoderover} extracts the abstract syntax trees in programs, then iteratively searches the useful information according to requirements and extracted ASTs, and eventually generates program patches. RepoUnderstander~\cite{ma2024understand} develops an exploration strategy based on the Monte Carlo Tree Search algorithm for software repository understanding and fault localization. While these agents have made significant strides in advancing ASE, it is important to note that the majority of these systems rely heavily on closed-source models. Open-source alternatives, while more accessible, have generally struggled to match the performance of their closed-source counterparts in complex software engineering tasks. Our work addresses this critical gap by developing an open-source model that aims to bridge the performance divide.

\subsection{Evaluation of Real-world Software Engineering Tasks} \label{evaluation_related_work}

Benefiting from the strong general capability of LLMs, LLM-based software engineering agents can handle many important SE tasks, e.g., code generation~\citep{wang2024codeact} and code debugging~\citep{hong2023metagpt}. More recently, SWE-bench team\citep{jimenez2023swe, yang2024sweagent} develop a unified dataset named SWE-bench to evaluate the capability of the agent system to resolve real-world GitHub issues automatically. Specifically, it collects the task instances from real-world GitHub issues from twelve repositories. Consistent with previous evaluation methods, SWE-bench is based on the automatic execution of the unit tests. Differently, the presented test set is challenging and requires the agents to have multiple capabilities simultaneously, including repository navigation, fault locating, debugging, code generation and program repairing, so as to solve a given issue end-to-end. Besides, SWE-bench Lite~\citep{swebenchlite} and SWE-bench Verified~\cite{swebenchverified} are subsets of SWE-bench, and they have a similar diversity and distribution of repositories as the original version. Due to the smaller test cost and more detailed human filtering, SWE-bench Lite and Verified are officially recommended as the benchmark of LLM-based SE agents. Therefore, consistent with previous methods~\citep{yang2024sweagent,autocoderover,xia2024agentless}, we report our performance on SWE-bench Lite and SWE-bench Verified.

\section{Lingma SWE-GPT}

\begin{figure*}
    \centering
    \includegraphics[scale=0.51]{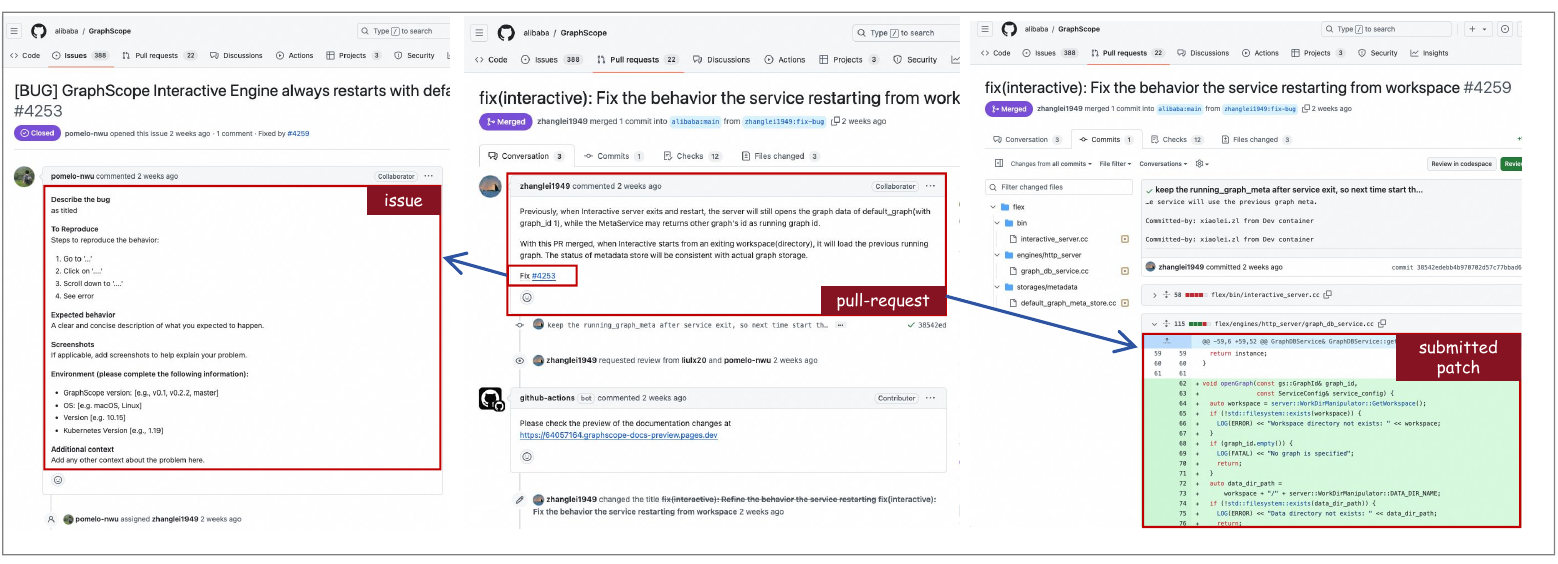}
    \caption{Example of issue and corresponding pull-request data collection process.}
    \label{fig:issue_pr}
\end{figure*}

In this section, we present our novel approach for training LLM to perform automated program improvement. Our method comprises three main phases: \textbf{issue data collection} (Figure \ref{fig:issue_pr}), \textbf{development process data synthesis} (Figure \ref{fig:process_graph}) and \textbf{model training} (Figure \ref{fig:training}). Leveraging an instruction-tuned model as the foundation model, our approach begins with the input of an issue description and the associated project codebase, then engages in a multi-stage workflow mimicking expert programmers' cognitive processes. This workflow consists of three key stages: repository understanding, fault localization, and patch generation. In the repository understanding stage, the model analyzes the repository structure, navigates the codebase, and utilizes a code compressor to grasp relevant project files and code snippets. The fault localization stage builds upon this understanding to identify potential fault locations at the class, function, or code level. Finally, in the patch generation stage, the model generates and applies patches using git-related operations. Throughout these stages, the model operates in a Chain-of-Thought ($CoT$) manner, where each step outputs a reasoning process ($CoT_{i}$) and an action ($Action_{i}$). This action is then executed, generating an observation ($Obs_{i}$). Based on this observation, the model proceeds to the next step ($CoT_{i+1}$ + $Action_{i+1}$), creating an iterative feedback loop for continuous refinement.


\subsection{Issue Data Collection}

The foundation of our approach lies in leveraging high-quality, real-world software development data. GitHub Issues serve as valuable resources for understanding bugs, feature requests, and enhancements, often providing natural language descriptions of problems or desired functionalities. Pull Requests (PRs) contain the corresponding code changes, including commit histories and code diffs, which directly address the issues raised. By leveraging this combination of data, we can simulate real-world programming scenarios, capturing the context and reasoning behind code modifications. To effectively train SWE-GPT for automatic program improvement, we constructed a comprehensive dataset by collecting issues, corresponding PRs, and codebases from public GitHub repositories. Below we will describe our data collection process in detail.

First, we selected Github repositories with at least 50 stars on GitHub, thereby ensuring a basic level of community recognition and activity. To avoid potential data leakage, we carefully filtered out any repositories that overlapped with those used in SWE-bench~\cite{jimenez2023swe}. For each selected repository, we retrieved all issues and their linked PRs, focusing specifically on those PRs that had been merged by the developers (as illustrated in Figure~\ref{fig:issue_pr}). Specifically, we utilized GitHub's API to fetch PRs with the state "merged" and their associated issues with the state "closed", ensuring that we captured only completed and approved code changes. Additionally, we stored snapshots of the codebase at the time of the PR to provide sufficient context for the code changes.

To further ensure the quality of the issues and PRs, we applied a set of heuristic filtering rules, similar to the approach used in OctoPack~\cite{muennighoff2023octopack}. \textbf{For issues}, we retained only those with textual descriptions containing at least 20 characters, thus excluding trivial or insufficiently detailed issues. Additionally, to avoid issues that primarily reference external resources without providing adequate context, we filtered out those with more than three hyperlinks. Lastly, we retained only issues with at least 80\% English content in their textual descriptions, to maintain language consistency across the dataset. \textbf{For PRs}, we applied two main criteria. First, we selected PRs that involved modifications to between one and five code files. This ensured that each PR represented a substantive but not overly complex change. Second, we excluded PRs that only modified test files, focusing instead on changes that directly impacted the primary functionality of the codebase.

After applying these filtering criteria, we constructed a dataset consisting of approximately 90,000 PRs collected from 4,000 repositories. Notably, the codebases we processed were typically of substantial size, often containing hundreds of files, reflecting the complexity inherent in real-world software improvements.

\subsection{Development Process Data Synthesis}

\begin{figure*}
    \centering
    \includegraphics[scale=0.41]{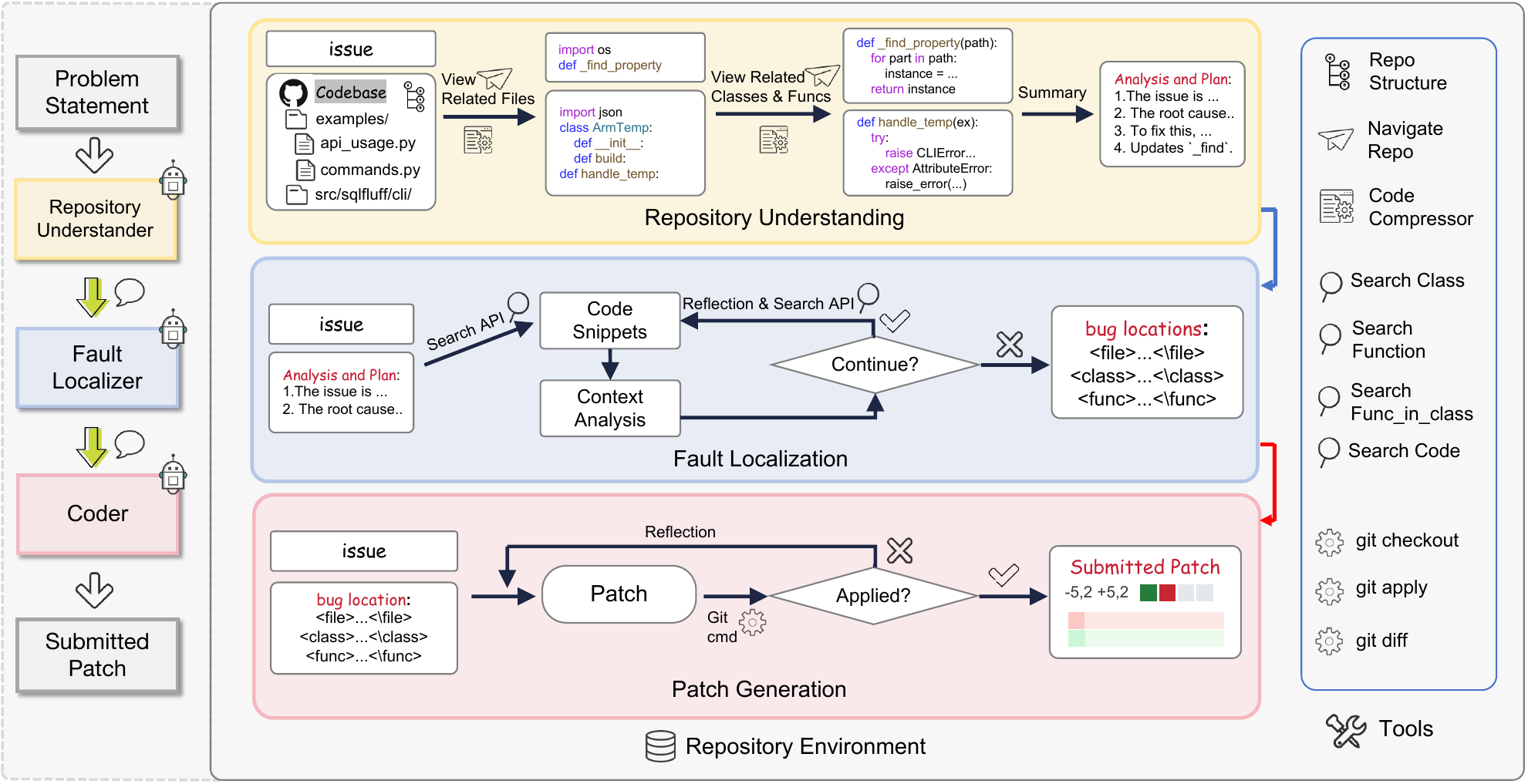}
    \caption{The overview of three-stage data synthesis and inference workflow.}
    \label{fig:process_graph}
\end{figure*}

Building upon the comprehensive issue data collection, we introduce a novel data synthesis process designed to capture the dynamic nature of software development. This process addresses the limitations of current LLM-based approaches by simulating the complex, interactive aspects of real-world software maintenance and evolution. Figure \ref{fig:training} illustrates our data synthesis approach, with its algorithm detailed in Algorithm \ref{algo:training}. This approach mimics the cognitive workflow of expert programmers across three key stages: Repository Understanding, Fault Localization, and Patch Generation. Each stage involves a series of Chain-of-Thought reasoning steps, allowing the model to iteratively refine its understanding and output to the given issue.

Figure \ref{fig:process_graph} illustrates our three-stage \underline{S}oft\underline{W}are \underline{E}ngineering process data \underline{Syn}thesis and \underline{Infer}ence workflow(line \ref{algo:line9}-\ref{algo:line18}, Algo. \ref{algo:training}), named SWESynInfer. This workflow extends the publicly available AutoCodeRover~\cite{autocoderover} framework. AutoCodeRover provides baseline processes for context retrieval and patch generation stages, our work further introduces crucial enhancements to more accurately simulate the cognitive processes of expert developers.

\textbf{Repository Understanding.} This stage focuses on comprehending the structure and content of a code repository, crucial for establishing the context necessary for issue resolution. The Repository Understanding Agent (RepoUer) employs a hierarchical approach inspired by Agentless \cite{xia2024agentless}, utilizing three key tools: Repo Structure tool, Navigate Repo tool, and Code Compressor tool (see Figure \ref{fig:process_graph}). First, RepoUer uses the Repo Structure tool to create a concise representation of the repository's directory structure. This tool transforms the directory structure into a tree format suitable for LLM input, starting from the repository's root folder and organizing code files and folder names. Files and folders at the same directory level are vertically aligned, with subdirectory contents indented. Next, the Navigate Repo tool is employed to traverse the repository and locate potentially relevant files and elements. RepoUer leverages the issue description and the generated directory tree to identify N potentially relevant files. The Code Compressor tool then transforms these files into a skeleton format. This compressed representation preserves global references, class definitions, function signatures, associated comments, and variable declarations, effectively reducing the overall context length while maintaining essential structural information. Using these compressed file representations, RepoUer further refines its search to identify specific classes or functions that are potentially relevant to the issue. Finally, RepoUer analyzes the collected information and formulates a comprehensive plan for issue resolution. This plan typically includes: a restatement of the issue, an analysis of the root cause, and proposed steps to fix the issue. This multi-step process emulates a developer's initial exploration of a project, providing a comprehensive understanding of the codebase's architecture and relevant components.

\textbf{Fault Localization.}
Building on insights from the repository understanding stage, the Fault Localization phase aims to identify specific problem areas within the codebase. This phase employs the Fault Localizer (FLer), which simulates the diagnostic steps developers take when resolving issues, combining tool utlization with iterative problem-solving. FLer initially uses specialized search APIs to retrieve relevant code snippets at the file, class, function, and snippet levels. These APIs (see Tools in Figure \ref{fig:process_graph}), informed by the repository summary and issue description, extract contextual information necessary for pinpointing potential fault locations. FLer then systematically evaluates the retrieved code snippets, performing context analysis to understand the relationships and dependencies within the codebase. Finally, it assesses whether the fault location is identified and decides whether to continue searching or proceed to the patch generation phase. FLer continues this cycle until it either successfully identifies the fault locations or reaches a iteration limit. FLer ensures that after each observation of results, it performs a summary and reflection. By documenting these intermediate reasoning steps, the model learns to emulate the cognitive processes involved in fault localization, enhancing its ability to tackle complex software engineering tasks.

\begin{algorithm}
\footnotesize
\caption{Lingma SWE-GPT Training Algorithm}
\label{algo:training}
\KwIn{Instruct LLM $M$, Dataset $D = \{(\text{issue}, \text{codebase}, \text{pull\_request})\}$, Fault localization threshold $m_1$, Patch similarity threshold $m_2$, Number of iterations $N$}
\KwOut{Optimized model $M'$}
Sort $D$ by pull-request timestamps to prevent data leakage\;
$M' \leftarrow M$  \greencomment{Initialize the model to be optimized}
\For{iteration $\leftarrow 1$ \KwTo $N$}{
    $B \leftarrow \{\}$  \greencomment{Initialize batch for current iteration}
    
    \ForEach{$(\text{issue}, \text{codebase}, \text{pull\_request}) \in D$}{
        submitted\_patch $\leftarrow$ extract\_patch(pull\_request)\;
        actual\_fault\_location $\leftarrow$ extract\_fault\_location(submitted\_patch)\; \label{algo:line7}
        
        Obs\_CoT\_Actions $\leftarrow$ []\;
        \ForEach{stage $\in$ \{\text{"Repo\_Understanding"}, \text{"Fault\_Localization"}, \text{"Patch\_Generation"}\}}{\label{algo:line9}
            stage\_Obs\_CoT\_Actions $\leftarrow$ []\;
            observation $\leftarrow$ initial\_observation(issue, codebase, stage, Obs\_CoT\_Action)\;
            \textcolor{green!50!black}{\tcp{Model performs autonomous analysis and reflection}}
            \While{not stage\_completed(stage, observation)}{ 
                cot\_action $\leftarrow$ generate\_cot\_action($M'$, observation, stage)\;
                Append (observation, cot\_action) to stage\_Obs\_CoT\_Actions\;
                next\_observation $\leftarrow$ execute\_action(cot\_action, codebase, stage)\;
                observation $\leftarrow$ next\_observation\;
            }
            Extend Obs\_CoT\_Actions with stage\_Obs\_CoT\_Actions\; \label{algo:line18}
        }
        \textcolor{green!50!black}{\tcp{Synthetic data filtering and selection}}
        predicted\_fault\_location $\leftarrow$ extract\_predicted\_location(Obs\_CoT\_Actions)\; \label{algo:line20}
        \If{similarity(predicted\_fault\_location, actual\_fault\_location) $\geq m_1$}{ \label{algo:line21}
            predicted\_patch $\leftarrow$ extract\_predicted\_patch(Obs\_CoT\_Actions)\;
            \If{patch\_similarity(predicted\_patch, submitted\_patch) $\geq m_2$}{  \label{algo:line23}
                $B \leftarrow B \cup \{(\text{issue}, \text{Obs\_CoT\_Actions})\}$\;
            }
            \Else{
                Obs\_CoT\_Actions\_without\_patch $\leftarrow$ remove\_patch\_related\_steps(Obs\_CoT\_Actions)\;
        $B \leftarrow B \cup \{(\text{issue}, \text{Obs\_CoT\_Actions\_without\_patch})\}$\; \label{algo:line27}

            }
        }
    }
    
    \textcolor{green!50!black}{\tcp{Optimize model using maximum likelihood estimation}}
    $\theta' \leftarrow \argmax_\theta \sum_{(\text{issue}, \text{Obs\_CoT\_Actions}) \in B} \sum_{(obs_i, cot\_action_i) \in \text{Obs\_CoT\_Actions}} \log P_\theta(cot\_action_i | \text{issue}, obs_i)$\; \label{algo:line29}
    $M' \leftarrow$ update\_model\_parameters($M$, $\theta'$)\; \label{algo:line30}
}
\Return{$M'$}
\end{algorithm}

\textbf{Patch Generation.}
In the final stage, the Patch Generation Agent (Coder) generates and applies patches to address the localized issues. This process involves patch generation and the use of git-related operations and lint tool to implement and validate changes. Specifically, Coder first generates a concrete solution based on the issue description and identified fault code snippets. It then replaces the fault code snippets with the new solution. If the generated patch fails to conform to the specified format or cannot be syntactically applied to the original program (i.e., if lint tools detect syntax errors in the generated code, or if the git apply command fails), Coder debugs based on the error type until a correctly formatted patch is generated or the maximum retry limit is reached. This stage embodies the iterative nature of real-world software development scenarios, allowing for multiple refinement cycles to produce high-quality, applicable patches.

By enhancing each stage with detailed intermediate reasoning steps and incorporating tools that mirror real-world development practices, SWESynInfer provides a more comprehensive and realistic simulation of the software maintenance and evolution process. This approach enables the generation of high-quality training data that captures the complex, interactive aspects of software development.

\textbf{Rejection Sampling.} To ensure the quality of the synthesized data, we implement a rejection sampling process based on two key metrics. This approach allows us to selectively retain high-quality instances that closely mimic real-world software development practices.

\textit{{Fault localization accuracy}}. We compare the predicted fault location with the actual fault location using a similarity threshold $m_{1}$. Specifically, we first extract the modified locations from the submitted patch in the pull request as the actual fault location (line \ref{algo:line7}, Algo. \ref{algo:training}). We map the patch to the current version of the repository and then use the abstract syntax tree to extract the corresponding functions and classes for the modified locations. For global code modifications, we select the surrounding code (3 lines above and below the modified line) as the modification location. We then use the same method to extract the modification location from the model-generated patch (line \ref{algo:line20}). To quantify the accuracy of fault localization, we calculate the Jaccard similarity coefficient~\cite{thada2013comparison} between these two sets of modification locations. Specifically, we divide the size of the intersection of the two sets by the size of their union. This ratio is then compared with the threshold $m_{1}$ (line \ref{algo:line21}, Algo. \ref{algo:training}). If the calculated ratio exceeds $m_{1}$, we consider the fault localization to be sufficiently accurate.

\textit{{Patch similarity}}. We evaluate the similarity between the predicted patch and the developer submitted patch using a threshold $m_{2}$. Following the approach in Agentless~\cite{xia2024agentless}, we first normalize both the model-generated and developer-written patches to ignore surface-level differences (e.g., extra spaces, newlines, and comments). We then calculate the similarity between the model-generated patch and the developer-written patch using both n-gram~\cite{cavnar1994ngram} and CodeBLEU~\cite{ren2020codebleu} scores (line \ref{algo:line23}, Algo. \ref{algo:training}). If any similarity score exceeds $m_{2}$, we retain the patch.

Data instances meeting both criteria are added to the training batch $B$ (line \ref{algo:line27}, Algo. \ref{algo:training}). For instances that accurately localize the fault but generate dissimilar patches, we retain the fault localization steps while removing patch-related actions, preserving valuable intermediate reasoning. These rigorous filtering criteria ensure that our synthesized data closely resembles real-world software development practices, also guaranteeing the reliability of the intermediate reasoning process.

\subsection{Model training}

After collecting a set of training examples $B_{i}$, we implement an iterative optimization strategy to train our model. In each iteration, the model optimizes its performance by maximizing the conditional probability of generating the target CoT and corresponding action given the current state observation (line \ref{algo:line29}, Algo. \ref{algo:training}). To further enhance the robustness of the training process, we incorporate a curriculum learning approach. As iterations progress, we accumulate problems that the current version of the model fails to solve and incrementally incorporate them into subsequent training. The complexity of the accumulated training examples increases progressively with the number of model iterations. This approach enables the model to establish a robust foundation on simpler tasks before addressing more complex challenges. Drawing inspiration from STaR~\cite{zelikman2022star} and NExT~\cite{ni2024next}, we adopt a similar strategy to mitigate the potential negative impact of low-quality samples that may exist in early iterations. Specifically, we initialize the model from its original checkpoint $M$ at the beginning of each iteration (line \ref{algo:line30}). This approach mitigates the risk of overfitting to potentially low-quality training examples in the early stages, thereby ensuring the stability of the training process and enhancing the generalization capability of the final model.

\section{Experiment Setup}

To evaluate the capabilities of Lingma SWE-GPT in resolving real-world Github issues, we answer the following research questions.

\textbf{RQ1:} How does Lingma SWE-GPT compare to state-of-the-art models in solving real-world software issues?

\textbf{RQ2:} What is the performance of Lingma SWE-GPT compared to open-source models in automated program improvement task?

\textbf{RQ3:} How effective is Lingma SWE-GPT in fault localization within the essential steps required for issue resolution?

\textbf{RQ4:} To what extent does the inherent randomness of large language models impact the consistency of Lingma SWE-GPT's performance?

\subsection{Benchmark and Evaluation Metric}
\textit{SWE-bench Verified and SWE-bench Lite.} We evaluated Lingma SWE-GPT on the recently proposed benchmarks SWE-bench Verified~\cite{swebenchverified} and SWE-bench Lite~\cite{swebenchlite}, comprising 500 and 300 real-world GitHub issues, respectively. The model receives only the natural language description of the original GitHub issue and its corresponding code repository as input. These benchmarks employ developer-written unit tests to verify the correctness of model-generated patches, ensuring a rigorous assessment of the model's performance. For detailed information on SWE-bench Verified and SWE-bench Lite, refer to Section \ref{evaluation_related_work}.

\textit{Evaluation Metric.} We use (1) the percentage of resolved task instances, (2) average inference cost of requesting closed-source model API. These evaluation metrics represent overall effectiveness, and economic efficacy in resolving real-world GitHub issues. Due to the public accessibility of open-source models, we assigns their API calls costs to NULL (-). This approach disregards potential deployment costs, which is left to future research to more comprehensively evaluate various factors, including indirect expenses. In addition, to mitigate the natural randomness of LLM, we repeat our experiments three times. Following AutoCodeRover~\cite{autocoderover} and SWE-agent~\cite{yang2024sweagent}, we report the results with the SWE-GPT @3 annotations (i.e., pass@3 metric).

\subsection{Baselines}
To thoroughly evaluate the performance of Lingma SWE-GPT in resolving real-world GitHub issues, we compared our model against both state-of-the-art closed-source and open-source models. 

\textbf{Closed-Source Models.} We included leading closed-source models whose results are reported on the SWE-bench leaderboard~\cite{jimenez2023swe} and in recent related research~\cite{zhu2024deepseekcoder}. These models primarily consist of OpenAI's GPT and Anthropic's Claude series. For these closed-source models, we utilized the results reported by SWE-bench~\cite{swebenchlite, swebenchverified}.

\begin{itemize}

\item \textbf{GPT-4}~\cite{achiam2023gpt4} and \textbf{GPT-4o}~\cite{gpt4o}: GPT-4 represents one of the most advanced language models available, exhibiting strong capabilities in understanding and generating human-like text and code. GPT-4o is OpenAI's flagship model that can reason across audio, vision, and text in real time. 

\item \textbf{Claude 3.5 Sonnet}~\cite{claude3.5} and \textbf{Claude 3 Ops}~\cite{claude3}, developed by Anthropic. The Claude models are designed with a focus on alignment and safe AI practices, and have shown proficiency in various tasks, particularly in code understanding and generation.

\item  \textbf{Gemini-1.5 Pro}~\cite{team2023gemini}, developed by Google. Gemini-1.5 Pro is a state-of-the-art multimodal model that excels in long-context understanding. It can process up to one million tokens in a single prompt, allowing for unprecedented context analysis.

\end{itemize}

\textbf{Open-Source Models.} For open-source baselines, we selected the most advanced models from the Llama and Qwen series. We obtained the results by both using existing research reports~\cite{liu2024codexgraph} and deploying and running them ourselves.

\begin{itemize}

\item \textbf{Llama Series}~\cite{llama3.1}: We evaluated Llama 3.1 instruction-tuned models with 70B and 405B parameters, representing the most advanced and largest open-source models in Llama 3.1 series.

\item \textbf{Qwen Series}~\cite{qwen2}: We included Qwen2-72B-Instruct, Qwen2.5-72B-Instruct, and Qwen2.5-Coder 7B (maximum coder version~\cite{hui2024qwen2.5-coder}) models. This series demonstrated superior performance on the HuggingFace open-source comprehensive leaderboard~\cite{huggingfaceleaderboard}.

\end{itemize}

\subsection{Implementation Details}

We implemented Lingma SWE-GPT using Qwen2.5 72B instruct~\cite{qwen2} and Qwen2.5 Coder 7B~\cite{hui2024qwen2.5-coder} as the foundation LLMs for the 72B and 7B versions, respectively. All training was conducted on a cluster of 64 NVIDIA A100 GPUs running Ubuntu 20.04, with a global batch size of 512 throughout the training process. For inference, we employed temperature sampling with a temperature ($T$) of 0.3. The training process consisted of 90 iterations in total. To mitigate the risk of the model converging on low-quality synthetic data during the initial stages, we utilized GPT-4o~\cite{gpt4o} to generate the initial training data for the first 10 iterations before transitioning to model-generated data. To ensure training efficiency, we updated the model with synthesized data every 10 iterations. In Algorithm \ref{algo:training}, the parameters $m_{1}$ and $m_{2}$ were set to 0.6 and 0.5, respectively. For evaluation purposes, we configured the inference process with a temperature of 0.3 and a maximum token limit of 1024 to generate results for each round. 
During the Repository Understanding stage, the number of potentially relevant files (N) identified was set to 5. In the Fault Localization stage, the predefined iteration limit was set to 5. For the Patch Generation stage, the maximum retry limit was set to 3.
To ensure consistency and reproducibility, all tests were conducted using the official SWE-bench Docker environment provided by the SWE-bench team~\cite{jimenez2023swe}.

\section{Evaluation}

\subsection{RQ1: Overall Effectiveness in Resolving Real-life Software Issues}

\begin{table}
\renewcommand{\arraystretch}{1.0}
\centering
\scalebox{1.0}{
\begin{tabular}{llccc}
\hline
\textbf{Agent} & \textbf{LLM} & \textbf{Verified} & \textbf{Lite} & \textbf{API Cost}\\
\hline
RAG~\cite{jimenez2023swe} & \includegraphics[height=1em]{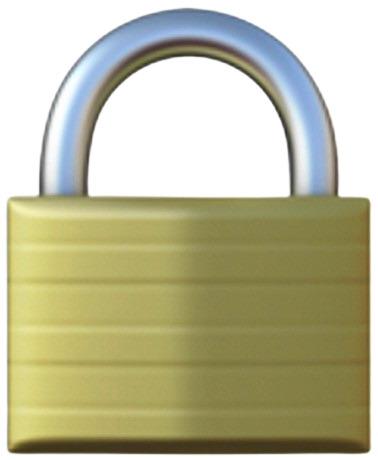} GPT-4  & 2.80\% & 2.67\% & \$0.13 \\
RAG~\cite{jimenez2023swe}	& \includegraphics[height=1em]{samples/imgs/lock-removebg-preview.png} Claude 3 Opus & 7.00\% & 4.33\% & \$0.25  \\
AutoCodeRover~\cite{liu2024codexgraph} & \includegraphics[height=1em]{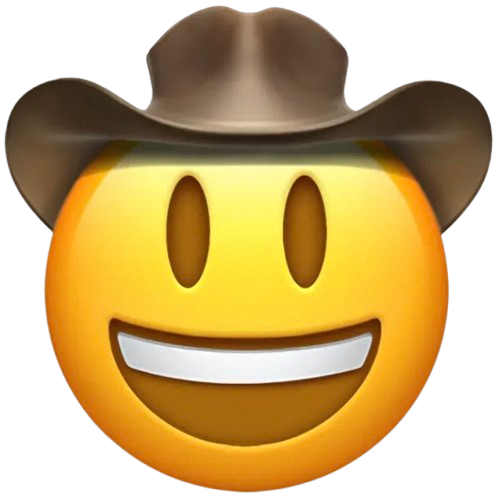} Qwen2 72B instruct & - & 9.34\% & \$ - \\

SWE-agent~\cite{jimenez2023swe} & \includegraphics[height=1em]{samples/imgs/lock-removebg-preview.png} Claude 3 Opus 	&  18.20\% & 11.67\% & \$3.42 \\
\rowcolor{gray!25}
SWESynInfer & \includegraphics[height=1em]{samples/imgs/smile-removebg-preview.png} Lingma SWE-GPT 7B & 18.20\% & 12.00\% & \$ - \\ 
SWE-agent~\cite{yang2024sweagent} & \includegraphics[height=1em]{samples/imgs/lock-removebg-preview.png} GPT-4 & 22.40\% & 18.00\% & \$2.51\\
SWE-agent~\cite{swebenchverified} & \includegraphics[height=1em]{samples/imgs/lock-removebg-preview.png} GPT-4o & 23.00\% &  18.30\% & \textit{Unknown} \\ 	
Refined OpenDevin~\cite{zhu2024deepseekcoder} & \includegraphics[height=1em]{samples/imgs/lock-removebg-preview.png} Gemini-1.5-Pro& - & 18.70\% & \textit{Unknown} \\

AutoCodeRover~\cite{autocoderover} & \includegraphics[height=1em]{samples/imgs/lock-removebg-preview.png} GPT-4 & - & 19.00\% & \$0.45\\
AppMap Navie~\cite{jimenez2023swe} & \includegraphics[height=1em]{samples/imgs/lock-removebg-preview.png} GPT-4o & 26.20\% & 21.67\% & \textit{Unknown} \\ 	
AutoCodeRover~\cite{swebenchverified} & \includegraphics[height=1em]{samples/imgs/lock-removebg-preview.png} GPT-4o & 28.80\% & 22.70\% & \textit{Unknown}\\
\rowcolor{gray!25}
SWESynInfer & \includegraphics[height=1em]{samples/imgs/smile-removebg-preview.png} Lingma SWE-GPT 72B & 30.20\% & 22.00\% & \$ - \\ 
SWESynInfer & \includegraphics[height=1em]{samples/imgs/lock-removebg-preview.png} GPT-4o & 31.80\% & 20.67\% & \$0.78 \\	
Agentless~\cite{swebenchverified} & \includegraphics[height=1em]{samples/imgs/lock-removebg-preview.png} GPT 4o & 33.20\% & \textbf{24.30\%} & \$0.34\\
SWE-agent~\cite{yang2024sweagent} & \includegraphics[height=1em]{samples/imgs/lock-removebg-preview.png} Claude 3.5 Sonnet & 33.60\% & 23.00\% & \textit{Unknown} \\ 	
\textbf{SWESynInfer} & \includegraphics[height=1em]{samples/imgs/lock-removebg-preview.png} \textbf{Claude 3.5 Sonnet} & \textbf{35.40\%} & 23.67\% & \$0.42 \\	\hline
\end{tabular}}
\caption{Performance comparison of Lingma SWE-GPT and other models on SWE-bench Verified and Lite benchmarks. Note: open-source models are denoted by \includegraphics[height=1em]{samples/imgs/smile-removebg-preview.png}, and closed-source models are indicated by \includegraphics[height=1em]{samples/imgs/lock-removebg-preview.png}. "-" signifies unreported results for the respective benchmark. "Unknown" indicates unreported API costs. For open-source models, "\$-" represents no direct API call costs.}
\label{tab:main_results}
\end{table}


To comprehensively evaluate the performance of Lingma SWE-GPT in resolving real-world software issues, we conducted extensive experiments using two challenging benchmarks: SWE-bench Verified and SWE-bench Lite. These benchmarks provide a realistic runtime testing environment that closely simulates real-world software development scenarios. The input for each task consists solely of the natural language description of GitHub issues and the corresponding repository code at the time the issue was reported. The expected output is a corrective patch that resolves the issue. We measure the overall effectiveness of Lingma SWE-GPT and baseline models by quantifying the number of successfully resolved issue instances. Table \ref{tab:main_results} presents the comprehensive results of Lingma SWE-GPT (7B and 72B) alongside various state-of-the-art models on both SWE-bench Verified and SWE-bench Lite. The results reveal several significant findings.

\textbf{Existing Open-Source vs. Closed-Source Models.} From table \ref{tab:main_results}, we can see that the majority of top-performing submissions are based on closed-source models. For instance, SWE-agent utilizing Claude 3.5 Sonnet achieves success rates of 33.60\% and 23.00\% on Verified and Lite benchmarks, respectively. Similarly, Agentless combined with GPT-4o resolves 33.20\% of issues on Verified and 24.30\% on Lite. These results underscore the current dominance of closed-source models in complex software engineering tasks. In contrast, open-source models have traditionally underperformed compared to their closed-source counterparts. For example, AutoCodeRover using Qwen2 72B instruct achieves a 9.34\% success rate on SWE-bench Lite, while the same framework with GPT-4o resolves 22.70\% of issues, highlighting a significant performance gap.

\textbf{{Performance of Lingma SWE-GPT.}} Lingma SWE-GPT 72B demonstrates competitive performance compared to state-of-the-art closed-source models. On SWE-bench Verified, it successfully resolves 30.20\% of the issues, closely approaching the performance of GPT-4o (31.80\%) under the same inference process. This marks a significant milestone as it is the first time an open-source model has surpassed the 30\% threshold in resolving these complex software issues. On SWE-bench Lite, Lingma SWE-GPT 72B even outperforms GPT-4o, resolving 22.00\% of issues compared to GPT-4o's 20.67\%. While Claude 3.5 Sonnet achieves the best overall performance on both benchmarks (35.40\% on Verified and 23.67\% on Lite), it's worth noting that Lingma SWE-GPT 72B's performance is also highly competitive, especially considering its open-source nature.

\textbf{{API Cost Considerations.}} A critical consideration in deploying large language models for software engineering tasks is the associated API costs. Due to the complexity of software repositories and the multi-round reasoning process, solving the 500 problems in SWE-bench Verified using GPT-4o incurs an approximate cost of \$390. This translates to an average of \$0.78 per issue, which can be prohibitively expensive for large-scale applications or continuous integration scenarios.

In contrast, open-source models like Lingma SWE-GPT do not incur direct API costs due to their public accessibility. This cost-effectiveness, combined with the competitive performance, presents a case for the adoption of open-source models in automated software engineering tasks. Interestingly, even the smaller Lingma SWE-GPT 7B model shows promising results, resolving 18.20\% and 12.00\% of issues in SWE-bench Verified and Lite, respectively. This performance underscores the potential of smaller models in automated issue resolution, particularly in resource-constrained scenarios. It opens up possibilities for more widespread integration of AI-assisted bug fixing and code improvement in software development pipelines, especially for organizations with budget constraints or privacy concerns.

\subsection{RQ2: Performance Against State-of-the-Art Open-Source Models}

\begin{table}
\renewcommand{\arraystretch}{1.0}
\centering
\scalebox{1.0}{
\begin{tabular}{llccc}
\hline
\textbf{Agent} & \textbf{LLM} &\textbf{Size}& \textbf{Verified} & \textbf{Lite}\\
\hline
SWESynInfer \textcolor{blue}{($\ast$)} & \includegraphics[height=1em]{samples/imgs/smile-removebg-preview.png} Qwen2.5-coder& 7B & 5.80\% & 4.33\% \\ 
SWESynInfer \textcolor{blue}{($\ast$)} & \includegraphics[height=1em]{samples/imgs/smile-removebg-preview.png} Llama-3.1-instruct & 70B & 17.20\% & 7.00\% \\
\rowcolor{gray!25}
SWESynInfer & \includegraphics[height=1em]{samples/imgs/smile-removebg-preview.png} Lingma SWE-GPT & 7B & 18.20\%  & 12.00\% \\
SWESynInfer & \includegraphics[height=1em]{samples/imgs/smile-removebg-preview.png} Qwen2-instruct  & 72B & 20.40\% & 13.70\% \\
SWESynInfer & \includegraphics[height=1em]{samples/imgs/smile-removebg-preview.png} Llama-3.1-instruct & 405B & 24.60\% & 15.66\% \\
SWESynInfer \textcolor{blue}{($\ast$)} & \includegraphics[height=1em]{samples/imgs/smile-removebg-preview.png} Qwen2.5-instruct & 72B & 25.40\% & 18.00\% \\

\rowcolor{gray!25}
\textbf{SWESynInfer} & \includegraphics[height=1em]{samples/imgs/smile-removebg-preview.png} \textbf{Lingma SWE-GPT} & \textbf{72B} & \textbf{30.20\% (18.90\%{\color{red} $\uparrow$})}  & \textbf{22.00\% (22.22\%{\color{red} $\uparrow$})} \\
\hline
\end{tabular}}
\caption{Comparative analysis of Lingma SWE-GPT and other open-source LLMs on SWE-bench Verified and Lite benchmarks. Note: \textcolor{blue}{($\ast$)} denotes models with limited instruction-following capabilities, necessitating manual prompt engineering, which may yield results that overestimate actual performance.}
\label{tab:open_source_comparison}
\end{table}

To evaluate the performance of Lingma SWE-GPT in comparison with state-of-the-art open-source models for automated program improvement, we conducted a comprehensive analysis using the same inference process (SWESynInfer) across various models. Table \ref{tab:open_source_comparison} presents the overall results of this comparison.

\textbf{{Experimental Challenges and Setup.}} During our evaluation, we encountered significant challenges with certain open-source models due to their limited instruction-following capabilities. To optimize the assessment of these models' effectiveness in software engineering tasks (rather than their general instruction-following ability), we implemented model-specific prompt engineering and tool customization. For instance, Qwen2.5-instruct 72B consistently generated fixed JSON formats $\verb|`|\verb|`|\verb|`|json\textbackslash n\{actual\_content\}\verb|`|\verb|`|\verb|`|$, while Llama-3.1-instruct 70B output $\verb|`|\verb|`|\verb|`|\textbackslash n\{actual\_content\}\verb|`|\verb|`|\verb|`|$, both of which led to JSON extraction failures(should be $\{actual\_content\}$). These issues likely originate from excessively constrained format requirements during the models' alignment processes, resulting in diminished adaptability to complex tasks. In contrast, larger models like Llama-3.1-instruct 405B and closed-source models demonstrated superior instruction-following abilities, correctly generating JSON formats as required. To ensure a fair comparison, we adjusted prompts and customized tool invocations for affected models, and indicated by \textcolor{blue}{($\ast$)} in Table \ref{tab:open_source_comparison}.

\textbf{Results on SWE-bench.} The results demonstrate that Lingma SWE-GPT 72B significantly outperforms the latest open-source models, including Qwen2.5-instruct 70B and the largest open-source model, Llama-3.1-instruct 405B. On SWE-bench Verified, Lingma SWE-GPT 72B achieves a 30.20\% success rate, marking an 18.90\% relative improvement over Qwen2.5-instruct 72B (25.40\%) and a 22.76\% relative improvement over Llama-3.1-instruct 405B (24.60\%). This performance gap highlights the effectiveness of Lingma SWE-GPT's training approach, which focuses on understanding and generating dynamically evolving processes in software development tasks.

Notably, even the smaller Lingma SWE-GPT 7B model outperforms Llama-3.1-instruct 70B (18.20\% vs. 17.20\% on SWE-bench Verified, 12.00\% vs. 7.00\% on SWE-bench Lite), further validating the efficacy of our process-oriented data training methodology. This result demonstrates that smaller, more efficient models can also attain competitive performance when trained on high-quality, process-oriented data.

Among other open-source models, Llama-3.1-instruct 405B exhibits the best performance without prompt-specific adjustments, highlighting its general capabilities and specialized abilities in software engineering tasks. This observation aligns with the scaling law hypothesis, which posits that larger models tend to perform better across a wide range of tasks. This insight provides valuable guidance for the future development of more advanced software engineering models.

\begin{figure*}
    \centering
    \includegraphics[scale=0.36]{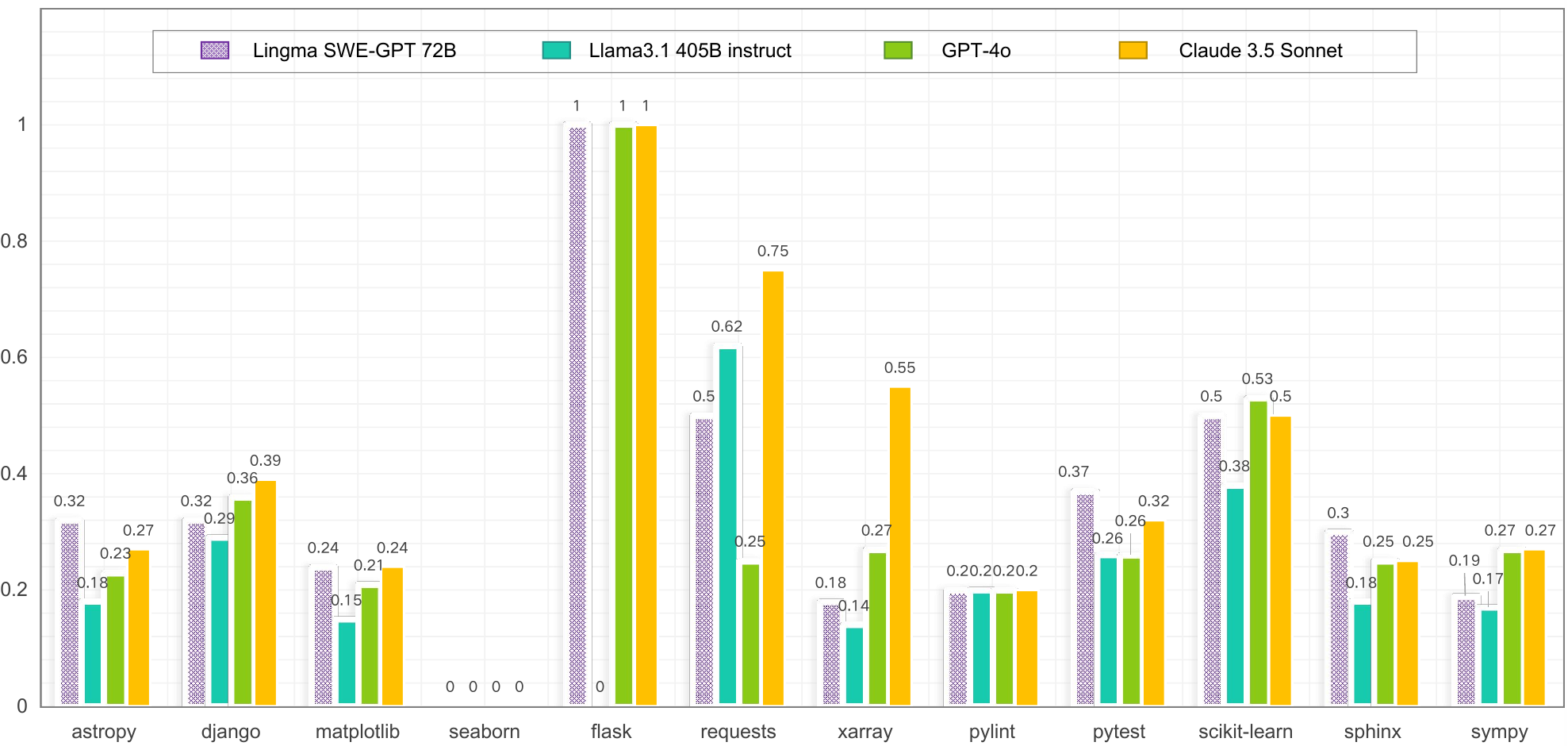}
    \caption{Comparison of issue resolution rates between Lingma SWE-GPT and other LLMs across different repositories in SWE-bench Verified.}
    \label{fig:resolved_by_repos}
\end{figure*}

\textbf{Performance Across Different Repositories.} Figure \ref{fig:resolved_by_repos} illustrates the performance of Lingma SWE-GPT 72B across 12 diverse software repositories, compared to the best-performing open-source model (Llama 3.1 405B instruct) and leading closed-source models (GPT-4o and Claude 3.5 Sonnet). Lingma SWE-GPT 72B outperforms Llama 3.1 405B instruct in the majority of repositories (9/12) and approaches the performance of closed-source models in numerous instances. This consistent performance across various domains demonstrates the model's robust generalization capabilities and its potential for extensive application in diverse software engineering contexts.

\begin{figure*}
    \centering
    \includegraphics[scale=0.99]{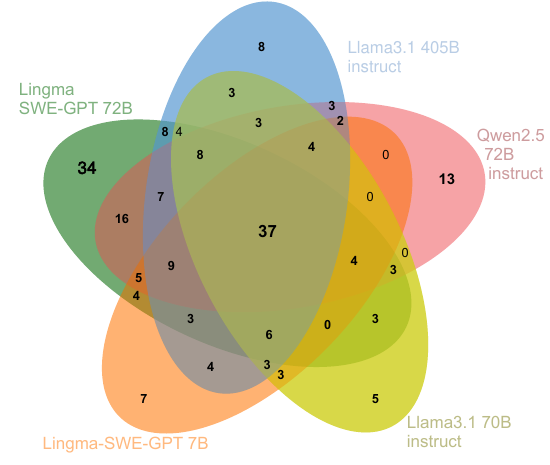}
    \caption{Venn diagram of issue instances solved by Lingma SWE-GPT and other open-source models on SWE-bench Verified.}
    \label{fig:jvenn}
\end{figure*}

\textbf{Complementarity of Models.} Figure \ref{fig:jvenn} presents a venn diagram illustrating the overlap and uniqueness of issues resolved by different models. Notably, Lingma SWE-GPT 72B uniquely resolves the highest number of issues (34), demonstrating its superior comprehension and analytical capabilities in software engineering processes. In addition, Lingma SWE-GPT 7B solves 7 tasks that are not solved by other models, surpassing the 5 independently solved tasks of Llama 3.1 70B-instruct, demonstrating the potential of smaller models. The diagram also reveals a degree of complementarity among different models, with each solving some unique issues. This observation suggests potential for future research in model ensemble approaches to further improve overall performance in automated program improvement tasks, a direction already began to explore in some existing studies~\cite{zhang2024diversity, aider}.

\subsection{RQ3: Fault Localization Effectiveness}

\begin{table}
\renewcommand{\arraystretch}{1.0}
\centering
\scalebox{1.0}{
\begin{tabular}{llcccc}
\hline
\textbf{LLM} &\textbf{Model Size}& \textbf{Verified} & \textbf{Chunk} & \textbf{Function} & \textbf{File}\\
\hline
\includegraphics[height=1em]{samples/imgs/smile-removebg-preview.png} Qwen2.5-coder-instruct& 7B & 5.80\% & 13.15\% & 13.91\% & 19.20\% \\ 
\includegraphics[height=1em]{samples/imgs/smile-removebg-preview.png} Llama-3.1-instruct & 70B & 17.20\% & 47.81\% & 51.01\% & 70.67\% \\
\rowcolor{gray!25}
\includegraphics[height=1em]{samples/imgs/smile-removebg-preview.png} Lingma SWE-GPT & 7B & 18.20\%  & 39.10\% & 42.20\% & 58.82\% \\
\includegraphics[height=1em]{samples/imgs/smile-removebg-preview.png} Qwen2-instruct  & 72B & 20.40\% & 38.87\% & 40.80\% & 55.21\% \\
\includegraphics[height=1em]{samples/imgs/smile-removebg-preview.png} Llama-3.1-instruct & 405B & 24.60\% & 44.92\%  & 47.80\% & 67.49\% \\ 
\includegraphics[height=1em]{samples/imgs/smile-removebg-preview.png} Qwen2.5-instruct & 72B & 25.40\% & 42.84\%  & 45.66\% & 61.34\%\\
\rowcolor{gray!25}
\includegraphics[height=1em]{samples/imgs/smile-removebg-preview.png} Lingma SWE-GPT ($run_1$) & 72B & 30.20\%  & 51.16\%  & 54.30\% & 72.29\% \\
\rowcolor{gray!25}
\includegraphics[height=1em]{samples/imgs/smile-removebg-preview.png} Lingma SWE-GPT ($run_2$) & 72B & 29.00\%  & 51.88\%  & 53.90\% & 72.02\% \\
\rowcolor{gray!25}
\includegraphics[height=1em]{samples/imgs/smile-removebg-preview.png} Lingma SWE-GPT ($run_3$) & 72B & 30.20\%  & 53.24\%  &  55.17\% &  72.90\% \\
\rowcolor{gray!25}
\includegraphics[height=1em]{samples/imgs/smile-removebg-preview.png} \textbf{Lingma SWE-GPT ($pass@3$)} & \textbf{72B} & \textbf{39.80\%}  & \textbf{61.63\%}  & \textbf{66.28\%} & \textbf{80.85\%} \\ 
\includegraphics[height=1em]{samples/imgs/lock-removebg-preview.png} GPT-4o & \textit{Unknown} & 31.80\% & 52.18\% & 55.68\% & 72.49\%\\	
\includegraphics[height=1em]{samples/imgs/lock-removebg-preview.png} \textbf{Claude 3.5 Sonnet} & \textbf{\textit{Unknown}} & \textbf{35.40\%} & \textbf{54.90\%} & \textbf{58.08\%} & \textbf{74.27\%} \\	
\hline
\end{tabular}}
\caption{Comparative analysis of fault localization accuracy across various language models at different granularities.}
\label{tab:bug_location}
\end{table}

Fault localization is a critical step in resolving software issues in real-world development scenarios. Accurate identification of edit locations is not only crucial for automated issue resolution and invaluable for assisting human developers in debugging tasks. To evaluate the fault localization capabilities of Lingma SWE-GPT, we conducted a comprehensive analysis comparing the locations of patches generated by various models to the actual patch locations. 

\textbf{Evaluation Methodology.} We employed a rigorous process to ensure a thorough evaluation. We mapped the patches to the current version of the repository and used abstract syntax trees to extract the functions and classes corresponding to the modified locations. For chunk-level analysis, which not only encompasses function/class modifications but also global code changes, we considered the surrounding code (3 lines above and below the modified line) as the global code modification location. Following the approach outlined in Agentless~\cite{xia2024agentless}, we acknowledge that while errors can be fixed at locations different from the actual patch, comparing with the real patch serves as an effective approximate measure. Table \ref{tab:bug_location} shows the localization accuracy at the chunk, function, and file levels with each model.

\textbf{Results and Analysis.} Our investigation reveals several key insights. Lingma SWE-GPT demonstrates significantly better fault localization capabilities compared to other open-source models across all granularity levels (chunk, function, and file). Its performance closely approaches that of closed-source models, underscoring its effectiveness in fault localization. In addition, we observe a general positive correlation between fault localization success rate and issue resolution rate. This relationship underscores the critical role of accurate fault localization in successful issue resolution. However, even the best-performing model, Claude 3.5 Sonnet, achieves localization accuracy of only 54.90\% and 58.08\% at chunk and function levels, respectively. This observation suggests that fault localization remains a challenging aspect of automated issue resolution. This presents an opportunity for future research to focus on improving localization techniques, potentially through enhanced code understanding mechanisms or more sophisticated static and dynamic analysis methods.

\subsection{RQ4: Model Consistency and Randomness Impact}

Large language models (LLMs) inherently exhibit stochastic behavior in their outputs, which can potentially affect their consistency and reliability in practical applications. To address this concern, we conducted a comprehensive investigation into the consistency of Lingma SWE-GPT's performance across multiple executions on the SWE-bench Verified. This analysis aims to quantify the impact of the model's inherent variability on its ability to consistently solve software engineering tasks.

\textbf{Evaluation Methodology.} We executed Lingma SWE-GPT 72B three times (denoted as run\textsubscript{1}, run\textsubscript{2}, run\textsubscript{3}) on the SWE-bench Verified dataset. This approach allows us to isolate and observe the effects of the model's inherent randomness on its performance. We then analyzed the results in terms of issue resolution rates and fault localization accuracy across different granularity levels (chunk, function, and file).

\textbf{Results and Analysis.} Table \ref{tab:bug_location} presents the detailed results of our consistency analysis. Across the three runs, Lingma SWE-GPT 72B demonstrated remarkable consistency in its issue resolution capability, achieving success rates of 30.20\%, 29.00\%, and 30.20\% for run\textsubscript{1}, run\textsubscript{2}, and run\textsubscript{3}, respectively. The fault localization performance also exhibited stability across runs, with narrow ranges at chunk, function, and file levels, further underscoring the model's consistency in pinpointing issue locations. Follow AutoCodeRover~\cite{autocoderover} and SWE-agent~\cite{jimenez2023swe}, we also implemented a pass@3 metric, which considers an issue as resolved if any of the three runs successfully addresses it. This approach significantly boosted the overall performance, increasing the issue resolution rate to 39.80\%. Notably, this surpasses the performance of Claude 3.5 Sonnet (35.40\%), a leading closed-source model. The pass@3 approach also substantially enhanced fault localization accuracy, achieving 61.63\% at the chunk level, 66.28\% at the function level, and 80.85\% at the file level. This is particularly relevant in scenarios where test suites or other validation mechanisms are available, opening avenues for more accurate automated software improvement techniques.

\section{Limitation and Threats to Validity}

While Lingma SWE-GPT has demonstrated promising results in automated software improvement, it is crucial to acknowledge the limitations of our approach and potential threats to the validity of our findings. This section discusses these aspects and outlines directions for future research.

\textbf{Automatic Solution Verification.} A limitation of our current approach is the lack of comprehensive solution verification step. Although Lingma SWE-GPT has shown impressive performance on the SWE-bench dataset, we have not implemented an automated process yet to verify the correctness of the generated patches through unit testing. This verification step is important for ensuring the reliability and practical applicability of the model's outputs. The primary challenge in addressing this limitation stems from the scarcity of suitable training data. Real-world scenarios rarely provide readily available datasets that encompass the entire process of testing and debugging. While it is possible to synthesize buggy code from existing repository data and simulate the patch verification and debugging process, this approach may not fully capture the complexity of real-world software issues. Furthermore, automatically constructing complex project environments and resolving dependencies poses a significant challenge. To address this, we propose developing an agent capable of automating environment setup, which we contend is crucial for scalable process data acquisition. We posit that future work will further enhance the end-to-end effectiveness of program improvement by incorporating solution verification mechanisms.

\textbf{Model Evaluation.} Another limitation of our study lies in the evaluation methodology. Our evaluation of Lingma SWE-GPT primarily utilizes the SWE-bench Verified~\cite{swebenchverified} and SWE-bench Lite~\cite{swebenchlite} benchmarks. While these benchmarks provide valuable insights into the model's ability to address real-world issues by assessing proficiency in various tasks such as fault localization, debugging, and program repair, they may not fully capture the complexity and real-world applicability of the model compared to human developers. We acknowledge that a more comprehensive evaluation approach would involve directly applying Lingma SWE-GPT to resolve open issues in open-source software projects on platforms like GitHub. This approach would offer a more direct validation of the model's efficacy in real-world software improvement scenarios. Furthermore, we are exploring more sophisticated evaluation methodologies to assess the model's performance on complex software engineering tasks that better reflect the nuances of real-world development processes.

Despite these limitations, Lingma SWE-GPT constitutes a significant step forward in automated software improvement. Each delineated limitation not only highlights the challenges in this field but also provides clear directions for future research and improvement. We aim to harness these findings to evolve Lingma SWE-GPT into a more robust, adaptable, and effective system for assisting developers throughout the software development lifecycle.

\section{Conclusion}

In this paper, we introduce Lingma SWE-GPT, a novel open-source large language model series designed to address complex software improvement tasks. This series includes two variants: Lingma SWE-GPT 7B and Lingma SWE-GPT 72B, catering to different computational resource requirements while maintaining high performance. By focusing on simulating the dynamic nature of software development, including tool usage reasoning and interactive problem-solving capabilities, Lingma SWE-GPT distinguishes itself apart from models trained solely on static code data. Our evaluation, utilizing the challenging SWE-bench Verified and Lite benchmarks demonstrates Lingma SWE-GPT's effectiveness in automated software improvement. The 72B variant consistently outperforms existing open-source models and approximates the performance of closed-source alternatives, achieving a remarkable 30.20\% success rate on SWE-bench Verified. Notably, the 7B variant resolves 18.20\% of issues, highlighting the potential of smaller models in resource-constrained environments. Lingma SWE-GPT achieves exceptional fault localization capabilities across chunk, function, and file levels. The model's consistent performance across multiple runs and the significant enhancements achieved through our pass@3 approach highlight its reliability and potential for ensemble methods in automated software engineering. As we continue to refine and expand its capabilities, we posit that Lingma SWE-GPT will play an increasingly significant role in supporting developers, enhancing productivity, and improving software quality, thereby advancing the field of AI-assisted software engineering.

\section{Data Availability}
We release our model checkpoints (Lingma SWE-GPT 7B \& 72B) and  data synthesis and inference (SWESynInfer) code to encourage further exploration in this direction. The artifact that supports the results discussed in this paper is available at: \url{https://github.com/LingmaTongyi/Lingma-SWE-GPT}

\begin{acks}
We would like to express our gratitude to Zhipeng Xue and Ke Liu for their invaluable feedback and suggestions on the manuscript.
\end{acks}

\bibliographystyle{ACM-Reference-Format}
\bibliography{reference}


\begin{thebibliography}{73}


\ifx \showCODEN    \undefined \def \showCODEN     #1{\unskip}     \fi
\ifx \showDOI      \undefined \def \showDOI       #1{#1}\fi
\ifx \showISBNx    \undefined \def \showISBNx     #1{\unskip}     \fi
\ifx \showISBNxiii \undefined \def \showISBNxiii  #1{\unskip}     \fi
\ifx \showISSN     \undefined \def \showISSN      #1{\unskip}     \fi
\ifx \showLCCN     \undefined \def \showLCCN      #1{\unskip}     \fi
\ifx \shownote     \undefined \def \shownote      #1{#1}          \fi
\ifx \showarticletitle \undefined \def \showarticletitle #1{#1}   \fi
\ifx \showURL      \undefined \def \showURL       {\relax}        \fi
\providecommand\bibfield[2]{#2}
\providecommand\bibinfo[2]{#2}
\providecommand\natexlab[1]{#1}
\providecommand\showeprint[2][]{arXiv:#2}

\bibitem[Achiam et~al\mbox{.}(2023)]%
        {achiam2023gpt4}
\bibfield{author}{\bibinfo{person}{Josh Achiam}, \bibinfo{person}{Steven Adler}, \bibinfo{person}{Sandhini Agarwal}, \bibinfo{person}{Lama Ahmad}, \bibinfo{person}{Ilge Akkaya}, \bibinfo{person}{Florencia~Leoni Aleman}, \bibinfo{person}{Diogo Almeida}, \bibinfo{person}{Janko Altenschmidt}, \bibinfo{person}{Sam Altman}, \bibinfo{person}{Shyamal Anadkat}, {et~al\mbox{.}}} \bibinfo{year}{2023}\natexlab{}.
\newblock \showarticletitle{Gpt-4 technical report}.
\newblock \bibinfo{journal}{\emph{arXiv preprint arXiv:2303.08774}} (\bibinfo{year}{2023}).
\newblock


\bibitem[AlOmar et~al\mbox{.}(2024)]%
        {alomar2024refactor}
\bibfield{author}{\bibinfo{person}{Eman~Abdullah AlOmar}, \bibinfo{person}{Anushkrishna Venkatakrishnan}, \bibinfo{person}{Mohamed~Wiem Mkaouer}, \bibinfo{person}{Christian Newman}, {and} \bibinfo{person}{Ali Ouni}.} \bibinfo{year}{2024}\natexlab{}.
\newblock \showarticletitle{How to refactor this code? An exploratory study on developer-ChatGPT refactoring conversations}. In \bibinfo{booktitle}{\emph{Proceedings of the 21st International Conference on Mining Software Repositories}}. \bibinfo{pages}{202--206}.
\newblock


\bibitem[{Anthropic}(2024a)]%
        {claude3.5}
\bibfield{author}{\bibinfo{person}{{Anthropic}}.} \bibinfo{year}{2024}\natexlab{a}.
\newblock \bibinfo{booktitle}{\emph{Introducing Claude 3.5 Sonnet}}.
\newblock
\urldef\tempurl%
\url{https://www.anthropic.com/news/claude-3-5-sonnet}
\showURL{%
\tempurl}


\bibitem[{Anthropic}(2024b)]%
        {claude3}
\bibfield{author}{\bibinfo{person}{{Anthropic}}.} \bibinfo{year}{2024}\natexlab{b}.
\newblock \bibinfo{booktitle}{\emph{Introducing the next generation of Claude}}.
\newblock
\urldef\tempurl%
\url{https://www.anthropic.com/news/claude-3-family}
\showURL{%
\tempurl}


\bibitem[{Carlos E. Jimenez, John Yang, Jiayi Geng}(2024)]%
        {swebenchlite}
\bibfield{author}{\bibinfo{person}{{Carlos E. Jimenez, John Yang, Jiayi Geng}}.} \bibinfo{year}{2024}\natexlab{}.
\newblock \bibinfo{booktitle}{\emph{SWE-bench Lite: A Canonical Subset for Efficient Evaluation of Language Models as Software Engineers}}.
\newblock
\urldef\tempurl%
\url{https://www.swebench.com/lite.html}
\showURL{%
\tempurl}


\bibitem[Cavnar et~al\mbox{.}(1994)]%
        {cavnar1994ngram}
\bibfield{author}{\bibinfo{person}{William~B Cavnar}, \bibinfo{person}{John~M Trenkle}, {et~al\mbox{.}}} \bibinfo{year}{1994}\natexlab{}.
\newblock \showarticletitle{N-gram-based text categorization}. In \bibinfo{booktitle}{\emph{Proceedings of SDAIR-94, 3rd annual symposium on document analysis and information retrieval}}, Vol.~\bibinfo{volume}{161175}. Ann Arbor, Michigan, \bibinfo{pages}{14}.
\newblock


\bibitem[Chakraborty and Ray(2021)]%
        {chakraborty2021multi}
\bibfield{author}{\bibinfo{person}{Saikat Chakraborty} {and} \bibinfo{person}{Baishakhi Ray}.} \bibinfo{year}{2021}\natexlab{}.
\newblock \showarticletitle{On multi-modal learning of editing source code}. In \bibinfo{booktitle}{\emph{2021 36th IEEE/ACM International Conference on Automated Software Engineering (ASE)}}. IEEE, \bibinfo{pages}{443--455}.
\newblock


\bibitem[Chen et~al\mbox{.}(2024)]%
        {chen2024coder}
\bibfield{author}{\bibinfo{person}{Dong Chen}, \bibinfo{person}{Shaoxin Lin}, \bibinfo{person}{Muhan Zeng}, \bibinfo{person}{Daoguang Zan}, \bibinfo{person}{Jian-Gang Wang}, \bibinfo{person}{Anton Cheshkov}, \bibinfo{person}{Jun Sun}, \bibinfo{person}{Hao Yu}, \bibinfo{person}{Guoliang Dong}, \bibinfo{person}{Artem Aliev}, {et~al\mbox{.}}} \bibinfo{year}{2024}\natexlab{}.
\newblock \showarticletitle{CodeR: Issue Resolving with Multi-Agent and Task Graphs}.
\newblock \bibinfo{journal}{\emph{arXiv preprint arXiv:2406.01304}} (\bibinfo{year}{2024}).
\newblock


\bibitem[Chen et~al\mbox{.}(2021)]%
        {humaneval}
\bibfield{author}{\bibinfo{person}{Mark Chen}, \bibinfo{person}{Jerry Tworek}, \bibinfo{person}{Heewoo Jun}, \bibinfo{person}{Qiming Yuan}, \bibinfo{person}{Henrique~Ponde de Oliveira~Pinto}, \bibinfo{person}{Jared Kaplan}, \bibinfo{person}{Harri Edwards}, \bibinfo{person}{Yuri Burda}, \bibinfo{person}{Nicholas Joseph}, \bibinfo{person}{Greg Brockman}, \bibinfo{person}{Alex Ray}, \bibinfo{person}{Raul Puri}, \bibinfo{person}{Gretchen Krueger}, \bibinfo{person}{Michael Petrov}, \bibinfo{person}{Heidy Khlaaf}, \bibinfo{person}{Girish Sastry}, \bibinfo{person}{Pamela Mishkin}, \bibinfo{person}{Brooke Chan}, \bibinfo{person}{Scott Gray}, \bibinfo{person}{Nick Ryder}, \bibinfo{person}{Mikhail Pavlov}, \bibinfo{person}{Alethea Power}, \bibinfo{person}{Lukasz Kaiser}, \bibinfo{person}{Mohammad Bavarian}, \bibinfo{person}{Clemens Winter}, \bibinfo{person}{Philippe Tillet}, \bibinfo{person}{Felipe~Petroski Such}, \bibinfo{person}{Dave Cummings}, \bibinfo{person}{Matthias Plappert}, \bibinfo{person}{Fotios Chantzis},
  \bibinfo{person}{Elizabeth Barnes}, \bibinfo{person}{Ariel Herbert-Voss}, \bibinfo{person}{William~Hebgen Guss}, \bibinfo{person}{Alex Nichol}, \bibinfo{person}{Alex Paino}, \bibinfo{person}{Nikolas Tezak}, \bibinfo{person}{Jie Tang}, \bibinfo{person}{Igor Babuschkin}, \bibinfo{person}{Suchir Balaji}, \bibinfo{person}{Shantanu Jain}, \bibinfo{person}{William Saunders}, \bibinfo{person}{Christopher Hesse}, \bibinfo{person}{Andrew~N. Carr}, \bibinfo{person}{Jan Leike}, \bibinfo{person}{Josh Achiam}, \bibinfo{person}{Vedant Misra}, \bibinfo{person}{Evan Morikawa}, \bibinfo{person}{Alec Radford}, \bibinfo{person}{Matthew Knight}, \bibinfo{person}{Miles Brundage}, \bibinfo{person}{Mira Murati}, \bibinfo{person}{Katie Mayer}, \bibinfo{person}{Peter Welinder}, \bibinfo{person}{Bob McGrew}, \bibinfo{person}{Dario Amodei}, \bibinfo{person}{Sam McCandlish}, \bibinfo{person}{Ilya Sutskever}, {and} \bibinfo{person}{Wojciech Zaremba}.} \bibinfo{year}{2021}\natexlab{}.
\newblock \showarticletitle{Evaluating Large Language Models Trained on Code}.
\newblock  (\bibinfo{year}{2021}).
\newblock
\showeprint[arxiv]{2107.03374}~[cs.LG]


\bibitem[Chen et~al\mbox{.}(2023)]%
        {chen2023teaching}
\bibfield{author}{\bibinfo{person}{Xinyun Chen}, \bibinfo{person}{Maxwell Lin}, \bibinfo{person}{Nathanael Sch{\"a}rli}, {and} \bibinfo{person}{Denny Zhou}.} \bibinfo{year}{2023}\natexlab{}.
\newblock \showarticletitle{Teaching large language models to self-debug}.
\newblock \bibinfo{journal}{\emph{arXiv preprint arXiv:2304.05128}} (\bibinfo{year}{2023}).
\newblock


\bibitem[{Cognition}(2023)]%
        {cognitionai2023devin}
\bibfield{author}{\bibinfo{person}{{Cognition}}.} \bibinfo{year}{2023}\natexlab{}.
\newblock \bibinfo{booktitle}{\emph{Introducing Devin}}.
\newblock
\urldef\tempurl%
\url{https://www.cognition.ai/introducing-devin}
\showURL{%
\tempurl}


\bibitem[Ding et~al\mbox{.}(2024)]%
        {ding2024cycle}
\bibfield{author}{\bibinfo{person}{Yangruibo Ding}, \bibinfo{person}{Marcus~J Min}, \bibinfo{person}{Gail Kaiser}, {and} \bibinfo{person}{Baishakhi Ray}.} \bibinfo{year}{2024}\natexlab{}.
\newblock \showarticletitle{Cycle: Learning to self-refine the code generation}.
\newblock \bibinfo{journal}{\emph{Proceedings of the ACM on Programming Languages}} \bibinfo{volume}{8}, \bibinfo{number}{OOPSLA1} (\bibinfo{year}{2024}), \bibinfo{pages}{392--418}.
\newblock


\bibitem[Gong et~al\mbox{.}(2024)]%
        {gong2024ast}
\bibfield{author}{\bibinfo{person}{Linyuan Gong}, \bibinfo{person}{Mostafa Elhoushi}, {and} \bibinfo{person}{Alvin Cheung}.} \bibinfo{year}{2024}\natexlab{}.
\newblock \showarticletitle{AST-T5: Structure-Aware Pretraining for Code Generation and Understanding}.
\newblock \bibinfo{journal}{\emph{arXiv preprint arXiv:2401.03003}} (\bibinfo{year}{2024}).
\newblock


\bibitem[Hong et~al\mbox{.}(2023)]%
        {hong2023metagpt}
\bibfield{author}{\bibinfo{person}{Sirui Hong}, \bibinfo{person}{Xiawu Zheng}, \bibinfo{person}{Jonathan Chen}, \bibinfo{person}{Yuheng Cheng}, \bibinfo{person}{Jinlin Wang}, \bibinfo{person}{Ceyao Zhang}, \bibinfo{person}{Zili Wang}, \bibinfo{person}{Steven Ka~Shing Yau}, \bibinfo{person}{Zijuan Lin}, \bibinfo{person}{Liyang Zhou}, {et~al\mbox{.}}} \bibinfo{year}{2023}\natexlab{}.
\newblock \showarticletitle{Metagpt: Meta programming for multi-agent collaborative framework}.
\newblock \bibinfo{journal}{\emph{arXiv preprint arXiv:2308.00352}} (\bibinfo{year}{2023}).
\newblock


\bibitem[Huang et~al\mbox{.}(2023)]%
        {huang2023towards}
\bibfield{author}{\bibinfo{person}{Xiangbing Huang}, \bibinfo{person}{Yingwei Ma}, \bibinfo{person}{Haifang Zhou}, \bibinfo{person}{Zhijie Jiang}, \bibinfo{person}{Yuanliang Zhang}, \bibinfo{person}{Teng Wang}, {and} \bibinfo{person}{Shanshan Li}.} \bibinfo{year}{2023}\natexlab{}.
\newblock \showarticletitle{Towards Better Multilingual Code Search through Cross-Lingual Contrastive Learning}. In \bibinfo{booktitle}{\emph{Proceedings of the 14th Asia-Pacific Symposium on Internetware}}. \bibinfo{pages}{22--32}.
\newblock


\bibitem[{Huggingface Open LLM Leaderboard}(2024)]%
        {huggingfaceleaderboard}
\bibfield{author}{\bibinfo{person}{{Huggingface Open LLM Leaderboard}}.} \bibinfo{year}{2024}\natexlab{}.
\newblock \bibinfo{booktitle}{\emph{Dataset Card for Evaluation run of Qwen}}.
\newblock
\urldef\tempurl%
\url{https://huggingface.co/datasets/open-llm-leaderboard/Qwen__Qwen2-72B-details}
\showURL{%
\tempurl}


\bibitem[Hui et~al\mbox{.}(2024)]%
        {hui2024qwen2.5-coder}
\bibfield{author}{\bibinfo{person}{Binyuan Hui}, \bibinfo{person}{Jian Yang}, \bibinfo{person}{Zeyu Cui}, \bibinfo{person}{Jiaxi Yang}, \bibinfo{person}{Dayiheng Liu}, \bibinfo{person}{Lei Zhang}, \bibinfo{person}{Tianyu Liu}, \bibinfo{person}{Jiajun Zhang}, \bibinfo{person}{Bowen Yu}, \bibinfo{person}{Kai Dang}, {et~al\mbox{.}}} \bibinfo{year}{2024}\natexlab{}.
\newblock \showarticletitle{Qwen2. 5-coder technical report}.
\newblock \bibinfo{journal}{\emph{arXiv preprint arXiv:2409.12186}} (\bibinfo{year}{2024}).
\newblock


\bibitem[Jiang et~al\mbox{.}(2023)]%
        {jiang2023automatic}
\bibfield{author}{\bibinfo{person}{Zhijie Jiang}, \bibinfo{person}{Haixu Xiong}, \bibinfo{person}{Yingwei Ma}, \bibinfo{person}{Yao Zhang}, \bibinfo{person}{Yan Ding}, \bibinfo{person}{Yun Xiong}, {and} \bibinfo{person}{Shanshan Li}.} \bibinfo{year}{2023}\natexlab{}.
\newblock \showarticletitle{Automatic Code Annotation Generation Based on Heterogeneous Graph Structure}. In \bibinfo{booktitle}{\emph{2023 IEEE International Conference on Software Analysis, Evolution and Reengineering (SANER)}}. IEEE, \bibinfo{pages}{497--508}.
\newblock


\bibitem[Jimenez et~al\mbox{.}(2023)]%
        {jimenez2023swe}
\bibfield{author}{\bibinfo{person}{Carlos~E Jimenez}, \bibinfo{person}{John Yang}, \bibinfo{person}{Alexander Wettig}, \bibinfo{person}{Shunyu Yao}, \bibinfo{person}{Kexin Pei}, \bibinfo{person}{Ofir Press}, {and} \bibinfo{person}{Karthik Narasimhan}.} \bibinfo{year}{2023}\natexlab{}.
\newblock \showarticletitle{Swe-bench: Can language models resolve real-world github issues?}
\newblock \bibinfo{journal}{\emph{arXiv preprint arXiv:2310.06770}} (\bibinfo{year}{2023}).
\newblock


\bibitem[Kong et~al\mbox{.}(2024)]%
        {kong2024contrastrepair}
\bibfield{author}{\bibinfo{person}{Jiaolong Kong}, \bibinfo{person}{Mingfei Cheng}, \bibinfo{person}{Xiaofei Xie}, \bibinfo{person}{Shangqing Liu}, \bibinfo{person}{Xiaoning Du}, {and} \bibinfo{person}{Qi Guo}.} \bibinfo{year}{2024}\natexlab{}.
\newblock \showarticletitle{ContrastRepair: Enhancing Conversation-Based Automated Program Repair via Contrastive Test Case Pairs}.
\newblock \bibinfo{journal}{\emph{arXiv preprint arXiv:2403.01971}} (\bibinfo{year}{2024}).
\newblock


\bibitem[Lee et~al\mbox{.}(2024)]%
        {lee2024unified}
\bibfield{author}{\bibinfo{person}{Cheryl Lee}, \bibinfo{person}{Chunqiu~Steven Xia}, \bibinfo{person}{Jen-tse Huang}, \bibinfo{person}{Zhouruixin Zhu}, \bibinfo{person}{Lingming Zhang}, {and} \bibinfo{person}{Michael~R Lyu}.} \bibinfo{year}{2024}\natexlab{}.
\newblock \showarticletitle{A Unified Debugging Approach via LLM-Based Multi-Agent Synergy}.
\newblock \bibinfo{journal}{\emph{arXiv preprint arXiv:2404.17153}} (\bibinfo{year}{2024}).
\newblock


\bibitem[Li et~al\mbox{.}(2023a)]%
        {li2023two}
\bibfield{author}{\bibinfo{person}{Jiaying Li}, \bibinfo{person}{Yan Lei}, \bibinfo{person}{Shanshan Li}, \bibinfo{person}{Haifang Zhou}, \bibinfo{person}{Yue Yu}, \bibinfo{person}{Zhouyang Jia}, \bibinfo{person}{Yingwei Ma}, {and} \bibinfo{person}{Teng Wang}.} \bibinfo{year}{2023}\natexlab{a}.
\newblock \showarticletitle{A two-stage framework for ambiguous classification in software engineering}. In \bibinfo{booktitle}{\emph{2023 IEEE 34th International Symposium on Software Reliability Engineering (ISSRE)}}. IEEE, \bibinfo{pages}{275--286}.
\newblock


\bibitem[Li et~al\mbox{.}(2023b)]%
        {li2023codeeditor}
\bibfield{author}{\bibinfo{person}{Jia Li}, \bibinfo{person}{Ge Li}, \bibinfo{person}{Zhuo Li}, \bibinfo{person}{Zhi Jin}, \bibinfo{person}{Xing Hu}, \bibinfo{person}{Kechi Zhang}, {and} \bibinfo{person}{Zhiyi Fu}.} \bibinfo{year}{2023}\natexlab{b}.
\newblock \showarticletitle{Codeeditor: Learning to edit source code with pre-trained models}.
\newblock \bibinfo{journal}{\emph{ACM Transactions on Software Engineering and Methodology}} \bibinfo{volume}{32}, \bibinfo{number}{6} (\bibinfo{year}{2023}), \bibinfo{pages}{1--22}.
\newblock


\bibitem[Li et~al\mbox{.}(2024)]%
        {li2024dllens}
\bibfield{author}{\bibinfo{person}{Meiziniu Li}, \bibinfo{person}{Dongze Li}, \bibinfo{person}{Jianmeng Liu}, \bibinfo{person}{Jialun Cao}, \bibinfo{person}{Yongqiang Tian}, {and} \bibinfo{person}{Shing-Chi Cheung}.} \bibinfo{year}{2024}\natexlab{}.
\newblock \showarticletitle{DLLens: Testing Deep Learning Libraries via LLM-aided Synthesis}.
\newblock \bibinfo{journal}{\emph{arXiv preprint arXiv:2406.07944}} (\bibinfo{year}{2024}).
\newblock


\bibitem[Liu et~al\mbox{.}(2024a)]%
        {liu2024mftcoder}
\bibfield{author}{\bibinfo{person}{Bingchang Liu}, \bibinfo{person}{Chaoyu Chen}, \bibinfo{person}{Zi Gong}, \bibinfo{person}{Cong Liao}, \bibinfo{person}{Huan Wang}, \bibinfo{person}{Zhichao Lei}, \bibinfo{person}{Ming Liang}, \bibinfo{person}{Dajun Chen}, \bibinfo{person}{Min Shen}, \bibinfo{person}{Hailian Zhou}, {et~al\mbox{.}}} \bibinfo{year}{2024}\natexlab{a}.
\newblock \showarticletitle{Mftcoder: Boosting code llms with multitask fine-tuning}. In \bibinfo{booktitle}{\emph{Proceedings of the 30th ACM SIGKDD Conference on Knowledge Discovery and Data Mining}}. \bibinfo{pages}{5430--5441}.
\newblock


\bibitem[Liu et~al\mbox{.}(2024d)]%
        {liu2024large}
\bibfield{author}{\bibinfo{person}{Junwei Liu}, \bibinfo{person}{Kaixin Wang}, \bibinfo{person}{Yixuan Chen}, \bibinfo{person}{Xin Peng}, \bibinfo{person}{Zhenpeng Chen}, \bibinfo{person}{Lingming Zhang}, {and} \bibinfo{person}{Yiling Lou}.} \bibinfo{year}{2024}\natexlab{d}.
\newblock \showarticletitle{Large Language Model-Based Agents for Software Engineering: A Survey}.
\newblock \bibinfo{journal}{\emph{arXiv preprint arXiv:2409.02977}} (\bibinfo{year}{2024}).
\newblock


\bibitem[Liu et~al\mbox{.}(2024e)]%
        {liu2024your}
\bibfield{author}{\bibinfo{person}{Jiawei Liu}, \bibinfo{person}{Chunqiu~Steven Xia}, \bibinfo{person}{Yuyao Wang}, {and} \bibinfo{person}{Lingming Zhang}.} \bibinfo{year}{2024}\natexlab{e}.
\newblock \showarticletitle{Is your code generated by chatgpt really correct? rigorous evaluation of large language models for code generation}.
\newblock \bibinfo{journal}{\emph{Advances in Neural Information Processing Systems}}  \bibinfo{volume}{36} (\bibinfo{year}{2024}).
\newblock


\bibitem[Liu et~al\mbox{.}(2024c)]%
        {liu2024codexgraph}
\bibfield{author}{\bibinfo{person}{Xiangyan Liu}, \bibinfo{person}{Bo Lan}, \bibinfo{person}{Zhiyuan Hu}, \bibinfo{person}{Yang Liu}, \bibinfo{person}{Zhicheng Zhang}, \bibinfo{person}{Wenmeng Zhou}, \bibinfo{person}{Fei Wang}, {and} \bibinfo{person}{Michael Shieh}.} \bibinfo{year}{2024}\natexlab{c}.
\newblock \showarticletitle{CodexGraph: Bridging Large Language Models and Code Repositories via Code Graph Databases}.
\newblock \bibinfo{journal}{\emph{arXiv preprint arXiv:2408.03910}} (\bibinfo{year}{2024}).
\newblock


\bibitem[Liu et~al\mbox{.}(2024b)]%
        {liu2024marscode}
\bibfield{author}{\bibinfo{person}{Yizhou Liu}, \bibinfo{person}{Pengfei Gao}, \bibinfo{person}{Xinchen Wang}, \bibinfo{person}{Chao Peng}, {and} \bibinfo{person}{Zhao Zhang}.} \bibinfo{year}{2024}\natexlab{b}.
\newblock \showarticletitle{MarsCode Agent: AI-native Automated Bug Fixing}.
\newblock \bibinfo{journal}{\emph{arXiv preprint arXiv:2409.00899}} (\bibinfo{year}{2024}).
\newblock


\bibitem[Lozhkov et~al\mbox{.}(2024)]%
        {lozhkov2024starcoder}
\bibfield{author}{\bibinfo{person}{Anton Lozhkov}, \bibinfo{person}{Raymond Li}, \bibinfo{person}{Loubna~Ben Allal}, \bibinfo{person}{Federico Cassano}, \bibinfo{person}{Joel Lamy-Poirier}, \bibinfo{person}{Nouamane Tazi}, \bibinfo{person}{Ao Tang}, \bibinfo{person}{Dmytro Pykhtar}, \bibinfo{person}{Jiawei Liu}, \bibinfo{person}{Yuxiang Wei}, {et~al\mbox{.}}} \bibinfo{year}{2024}\natexlab{}.
\newblock \showarticletitle{StarCoder 2 and The Stack v2: The Next Generation}.
\newblock \bibinfo{journal}{\emph{arXiv preprint arXiv:2402.19173}} (\bibinfo{year}{2024}).
\newblock


\bibitem[Luo et~al\mbox{.}(2024)]%
        {luo2024repoagent}
\bibfield{author}{\bibinfo{person}{Qinyu Luo}, \bibinfo{person}{Yining Ye}, \bibinfo{person}{Shihao Liang}, \bibinfo{person}{Zhong Zhang}, \bibinfo{person}{Yujia Qin}, \bibinfo{person}{Yaxi Lu}, \bibinfo{person}{Yesai Wu}, \bibinfo{person}{Xin Cong}, \bibinfo{person}{Yankai Lin}, \bibinfo{person}{Yingli Zhang}, {et~al\mbox{.}}} \bibinfo{year}{2024}\natexlab{}.
\newblock \showarticletitle{RepoAgent: An LLM-Powered Open-Source Framework for Repository-level Code Documentation Generation}.
\newblock \bibinfo{journal}{\emph{arXiv preprint arXiv:2402.16667}} (\bibinfo{year}{2024}).
\newblock


\bibitem[Luo et~al\mbox{.}(2023)]%
        {luo2023wizardcoder}
\bibfield{author}{\bibinfo{person}{Ziyang Luo}, \bibinfo{person}{Can Xu}, \bibinfo{person}{Pu Zhao}, \bibinfo{person}{Qingfeng Sun}, \bibinfo{person}{Xiubo Geng}, \bibinfo{person}{Wenxiang Hu}, \bibinfo{person}{Chongyang Tao}, \bibinfo{person}{Jing Ma}, \bibinfo{person}{Qingwei Lin}, {and} \bibinfo{person}{Daxin Jiang}.} \bibinfo{year}{2023}\natexlab{}.
\newblock \showarticletitle{Wizardcoder: Empowering code large language models with evol-instruct}.
\newblock \bibinfo{journal}{\emph{arXiv preprint arXiv:2306.08568}} (\bibinfo{year}{2023}).
\newblock


\bibitem[Ma et~al\mbox{.}(2023a)]%
        {ma2023training}
\bibfield{author}{\bibinfo{person}{Yingwei Ma}, \bibinfo{person}{Yue Liu}, \bibinfo{person}{Yue Yu}, \bibinfo{person}{Yuanliang Zhang}, \bibinfo{person}{Yu Jiang}, \bibinfo{person}{Changjian Wang}, {and} \bibinfo{person}{Shanshan Li}.} \bibinfo{year}{2023}\natexlab{a}.
\newblock \showarticletitle{At Which Training Stage Does Code Data Help LLMs Reasoning?}
\newblock \bibinfo{journal}{\emph{arXiv preprint arXiv:2309.16298}} (\bibinfo{year}{2023}).
\newblock


\bibitem[Ma et~al\mbox{.}(2024)]%
        {ma2024understand}
\bibfield{author}{\bibinfo{person}{Yingwei Ma}, \bibinfo{person}{Qingping Yang}, \bibinfo{person}{Rongyu Cao}, \bibinfo{person}{Binhua Li}, \bibinfo{person}{Fei Huang}, {and} \bibinfo{person}{Yongbin Li}.} \bibinfo{year}{2024}\natexlab{}.
\newblock \showarticletitle{How to Understand Whole Software Repository?}
\newblock \bibinfo{journal}{\emph{arXiv preprint arXiv:2406.01422}} (\bibinfo{year}{2024}).
\newblock


\bibitem[Ma et~al\mbox{.}(2023b)]%
        {ma2023mulcs}
\bibfield{author}{\bibinfo{person}{Yingwei Ma}, \bibinfo{person}{Yue Yu}, \bibinfo{person}{Shanshan Li}, \bibinfo{person}{Zhouyang Jia}, \bibinfo{person}{Jun Ma}, \bibinfo{person}{Rulin Xu}, \bibinfo{person}{Wei Dong}, {and} \bibinfo{person}{Xiangke Liao}.} \bibinfo{year}{2023}\natexlab{b}.
\newblock \showarticletitle{Mulcs: Towards a unified deep representation for multilingual code search}. In \bibinfo{booktitle}{\emph{2023 IEEE International Conference on Software Analysis, Evolution and Reengineering (SANER)}}. IEEE, \bibinfo{pages}{120--131}.
\newblock


\bibitem[Meta(2024)]%
        {llama3.1}
\bibfield{author}{\bibinfo{person}{Meta}.} \bibinfo{year}{2024}\natexlab{}.
\newblock \bibinfo{booktitle}{\emph{Introducing Llama 3.1}}.
\newblock
\urldef\tempurl%
\url{https://ai.meta.com/blog/meta-llama-3-1/}
\showURL{%
\tempurl}


\bibitem[Muennighoff et~al\mbox{.}(2023)]%
        {muennighoff2023octopack}
\bibfield{author}{\bibinfo{person}{Niklas Muennighoff}, \bibinfo{person}{Qian Liu}, \bibinfo{person}{Armel Zebaze}, \bibinfo{person}{Qinkai Zheng}, \bibinfo{person}{Binyuan Hui}, \bibinfo{person}{Terry~Yue Zhuo}, \bibinfo{person}{Swayam Singh}, \bibinfo{person}{Xiangru Tang}, \bibinfo{person}{Leandro Von~Werra}, {and} \bibinfo{person}{Shayne Longpre}.} \bibinfo{year}{2023}\natexlab{}.
\newblock \showarticletitle{Octopack: Instruction tuning code large language models}.
\newblock \bibinfo{journal}{\emph{arXiv preprint arXiv:2308.07124}} (\bibinfo{year}{2023}).
\newblock


\bibitem[Ni et~al\mbox{.}(2024)]%
        {ni2024next}
\bibfield{author}{\bibinfo{person}{Ansong Ni}, \bibinfo{person}{Miltiadis Allamanis}, \bibinfo{person}{Arman Cohan}, \bibinfo{person}{Yinlin Deng}, \bibinfo{person}{Kensen Shi}, \bibinfo{person}{Charles Sutton}, {and} \bibinfo{person}{Pengcheng Yin}.} \bibinfo{year}{2024}\natexlab{}.
\newblock \showarticletitle{NExT: Teaching Large Language Models to Reason about Code Execution}.
\newblock \bibinfo{journal}{\emph{arXiv preprint arXiv:2404.14662}} (\bibinfo{year}{2024}).
\newblock


\bibitem[{OpenAI}(2024a)]%
        {gpt4o}
\bibfield{author}{\bibinfo{person}{{OpenAI}}.} \bibinfo{year}{2024}\natexlab{a}.
\newblock \bibinfo{booktitle}{\emph{Introducing GPT-4o}}.
\newblock
\urldef\tempurl%
\url{https://openai.com/index/hello-gpt-4o/}
\showURL{%
\tempurl}


\bibitem[{OpenAI}(2024b)]%
        {swebenchverified}
\bibfield{author}{\bibinfo{person}{{OpenAI}}.} \bibinfo{year}{2024}\natexlab{b}.
\newblock \bibinfo{booktitle}{\emph{Introducing SWE-bench Verified}}.
\newblock
\urldef\tempurl%
\url{https://openai.com/index/introducing-swe-bench-verified/}
\showURL{%
\tempurl}


\bibitem[Pan et~al\mbox{.}(2023)]%
        {pan2023understanding}
\bibfield{author}{\bibinfo{person}{Rangeet Pan}, \bibinfo{person}{Ali~Reza Ibrahimzada}, \bibinfo{person}{Rahul Krishna}, \bibinfo{person}{Divya Sankar}, \bibinfo{person}{Lambert~Pouguem Wassi}, \bibinfo{person}{Michele Merler}, \bibinfo{person}{Boris Sobolev}, \bibinfo{person}{Raju Pavuluri}, \bibinfo{person}{Saurabh Sinha}, {and} \bibinfo{person}{Reyhaneh Jabbarvand}.} \bibinfo{year}{2023}\natexlab{}.
\newblock \showarticletitle{Understanding the effectiveness of large language models in code translation}.
\newblock \bibinfo{journal}{\emph{arXiv preprint arXiv:2308.03109}} (\bibinfo{year}{2023}).
\newblock


\bibitem[Pan et~al\mbox{.}(2024)]%
        {pan2024codev}
\bibfield{author}{\bibinfo{person}{Zhenyu Pan}, \bibinfo{person}{Rongyu Cao}, \bibinfo{person}{Yongchang Cao}, \bibinfo{person}{Yingwei Ma}, \bibinfo{person}{Binhua Li}, \bibinfo{person}{Fei Huang}, \bibinfo{person}{Han Liu}, {and} \bibinfo{person}{Yongbin Li}.} \bibinfo{year}{2024}\natexlab{}.
\newblock \showarticletitle{Codev-Bench: How Do LLMs Understand Developer-Centric Code Completion?}
\newblock \bibinfo{journal}{\emph{arXiv preprint arXiv:2410.01353}} (\bibinfo{year}{2024}).
\newblock


\bibitem[{Paul Gauthier}(2024)]%
        {aider}
\bibfield{author}{\bibinfo{person}{{Paul Gauthier}}.} \bibinfo{year}{2024}\natexlab{}.
\newblock \bibinfo{booktitle}{\emph{Aider is ai pair programming in your terminal.}}
\newblock
\urldef\tempurl%
\url{https://aider.chat/2024}
\showURL{%
\tempurl}


\bibitem[Ren et~al\mbox{.}(2020)]%
        {ren2020codebleu}
\bibfield{author}{\bibinfo{person}{Shuo Ren}, \bibinfo{person}{Daya Guo}, \bibinfo{person}{Shuai Lu}, \bibinfo{person}{Long Zhou}, \bibinfo{person}{Shujie Liu}, \bibinfo{person}{Duyu Tang}, \bibinfo{person}{Neel Sundaresan}, \bibinfo{person}{Ming Zhou}, \bibinfo{person}{Ambrosio Blanco}, {and} \bibinfo{person}{Shuai Ma}.} \bibinfo{year}{2020}\natexlab{}.
\newblock \showarticletitle{Codebleu: a method for automatic evaluation of code synthesis}.
\newblock \bibinfo{journal}{\emph{arXiv preprint arXiv:2009.10297}} (\bibinfo{year}{2020}).
\newblock


\bibitem[Roziere et~al\mbox{.}(2023)]%
        {codellama}
\bibfield{author}{\bibinfo{person}{Baptiste Roziere}, \bibinfo{person}{Jonas Gehring}, \bibinfo{person}{Fabian Gloeckle}, \bibinfo{person}{Sten Sootla}, \bibinfo{person}{Itai Gat}, \bibinfo{person}{Xiaoqing~Ellen Tan}, \bibinfo{person}{Yossi Adi}, \bibinfo{person}{Jingyu Liu}, \bibinfo{person}{Romain Sauvestre}, \bibinfo{person}{Tal Remez}, {et~al\mbox{.}}} \bibinfo{year}{2023}\natexlab{}.
\newblock \showarticletitle{Code llama: Open foundation models for code}.
\newblock \bibinfo{journal}{\emph{arXiv preprint arXiv:2308.12950}} (\bibinfo{year}{2023}).
\newblock


\bibitem[Shen et~al\mbox{.}(2023)]%
        {shen2023pangu}
\bibfield{author}{\bibinfo{person}{Bo Shen}, \bibinfo{person}{Jiaxin Zhang}, \bibinfo{person}{Taihong Chen}, \bibinfo{person}{Daoguang Zan}, \bibinfo{person}{Bing Geng}, \bibinfo{person}{An Fu}, \bibinfo{person}{Muhan Zeng}, \bibinfo{person}{Ailun Yu}, \bibinfo{person}{Jichuan Ji}, \bibinfo{person}{Jingyang Zhao}, {et~al\mbox{.}}} \bibinfo{year}{2023}\natexlab{}.
\newblock \showarticletitle{Pangu-coder2: Boosting large language models for code with ranking feedback}.
\newblock \bibinfo{journal}{\emph{arXiv preprint arXiv:2307.14936}} (\bibinfo{year}{2023}).
\newblock


\bibitem[Shi et~al\mbox{.}(2024)]%
        {shi2024code}
\bibfield{author}{\bibinfo{person}{Yuling Shi}, \bibinfo{person}{Songsong Wang}, \bibinfo{person}{Chengcheng Wan}, {and} \bibinfo{person}{Xiaodong Gu}.} \bibinfo{year}{2024}\natexlab{}.
\newblock \showarticletitle{From Code to Correctness: Closing the Last Mile of Code Generation with Hierarchical Debugging}.
\newblock \bibinfo{journal}{\emph{arXiv preprint arXiv:2410.01215}} (\bibinfo{year}{2024}).
\newblock


\bibitem[Shirafuji et~al\mbox{.}(2023)]%
        {shirafuji2023refactoring}
\bibfield{author}{\bibinfo{person}{Atsushi Shirafuji}, \bibinfo{person}{Yusuke Oda}, \bibinfo{person}{Jun Suzuki}, \bibinfo{person}{Makoto Morishita}, {and} \bibinfo{person}{Yutaka Watanobe}.} \bibinfo{year}{2023}\natexlab{}.
\newblock \showarticletitle{Refactoring programs using large language models with few-shot examples}. In \bibinfo{booktitle}{\emph{2023 30th Asia-Pacific Software Engineering Conference (APSEC)}}. IEEE, \bibinfo{pages}{151--160}.
\newblock


\bibitem[Team(2024)]%
        {team2024codegemma}
\bibfield{author}{\bibinfo{person}{CodeGemma Team}.} \bibinfo{year}{2024}\natexlab{}.
\newblock \showarticletitle{Codegemma: Open code models based on gemma}.
\newblock \bibinfo{journal}{\emph{arXiv preprint arXiv:2406.11409}} (\bibinfo{year}{2024}).
\newblock


\bibitem[Team et~al\mbox{.}(2023)]%
        {team2023gemini}
\bibfield{author}{\bibinfo{person}{Gemini Team}, \bibinfo{person}{Rohan Anil}, \bibinfo{person}{Sebastian Borgeaud}, \bibinfo{person}{Yonghui Wu}, \bibinfo{person}{Jean-Baptiste Alayrac}, \bibinfo{person}{Jiahui Yu}, \bibinfo{person}{Radu Soricut}, \bibinfo{person}{Johan Schalkwyk}, \bibinfo{person}{Andrew~M Dai}, \bibinfo{person}{Anja Hauth}, {et~al\mbox{.}}} \bibinfo{year}{2023}\natexlab{}.
\newblock \showarticletitle{Gemini: a family of highly capable multimodal models}.
\newblock \bibinfo{journal}{\emph{arXiv preprint arXiv:2312.11805}} (\bibinfo{year}{2023}).
\newblock


\bibitem[Thada and Jaglan(2013)]%
        {thada2013comparison}
\bibfield{author}{\bibinfo{person}{Vikas Thada} {and} \bibinfo{person}{Vivek Jaglan}.} \bibinfo{year}{2013}\natexlab{}.
\newblock \showarticletitle{Comparison of jaccard, dice, cosine similarity coefficient to find best fitness value for web retrieved documents using genetic algorithm}.
\newblock \bibinfo{journal}{\emph{International Journal of Innovations in Engineering and Technology}} \bibinfo{volume}{2}, \bibinfo{number}{4} (\bibinfo{year}{2013}), \bibinfo{pages}{202--205}.
\newblock


\bibitem[Wang et~al\mbox{.}(2024)]%
        {wang2024codeact}
\bibfield{author}{\bibinfo{person}{Xingyao Wang}, \bibinfo{person}{Yangyi Chen}, \bibinfo{person}{Lifan Yuan}, \bibinfo{person}{Yizhe Zhang}, \bibinfo{person}{Yunzhu Li}, \bibinfo{person}{Hao Peng}, {and} \bibinfo{person}{Heng Ji}.} \bibinfo{year}{2024}\natexlab{}.
\newblock \bibinfo{title}{Executable Code Actions Elicit Better LLM Agents}.
\newblock
\newblock
\showeprint[arxiv]{2402.01030}~[cs.CL]


\bibitem[Wei et~al\mbox{.}(2024)]%
        {wei2024magicoder}
\bibfield{author}{\bibinfo{person}{Yuxiang Wei}, \bibinfo{person}{Zhe Wang}, \bibinfo{person}{Jiawei Liu}, \bibinfo{person}{Yifeng Ding}, {and} \bibinfo{person}{Lingming Zhang}.} \bibinfo{year}{2024}\natexlab{}.
\newblock \showarticletitle{Magicoder: Empowering code generation with oss-instruct}. In \bibinfo{booktitle}{\emph{Forty-first International Conference on Machine Learning}}.
\newblock


\bibitem[Xia et~al\mbox{.}(2024a)]%
        {xia2024agentless}
\bibfield{author}{\bibinfo{person}{Chunqiu~Steven Xia}, \bibinfo{person}{Yinlin Deng}, \bibinfo{person}{Soren Dunn}, {and} \bibinfo{person}{Lingming Zhang}.} \bibinfo{year}{2024}\natexlab{a}.
\newblock \showarticletitle{Agentless: Demystifying llm-based software engineering agents}.
\newblock \bibinfo{journal}{\emph{arXiv preprint arXiv:2407.01489}} (\bibinfo{year}{2024}).
\newblock


\bibitem[Xia et~al\mbox{.}(2024b)]%
        {xia2024fuzz4all}
\bibfield{author}{\bibinfo{person}{Chunqiu~Steven Xia}, \bibinfo{person}{Matteo Paltenghi}, \bibinfo{person}{Jia Le~Tian}, \bibinfo{person}{Michael Pradel}, {and} \bibinfo{person}{Lingming Zhang}.} \bibinfo{year}{2024}\natexlab{b}.
\newblock \showarticletitle{Fuzz4all: Universal fuzzing with large language models}. In \bibinfo{booktitle}{\emph{Proceedings of the IEEE/ACM 46th International Conference on Software Engineering}}. \bibinfo{pages}{1--13}.
\newblock


\bibitem[Xie et~al\mbox{.}(2024)]%
        {xie2024pet}
\bibfield{author}{\bibinfo{person}{Yifan Xie}, \bibinfo{person}{Zhouyang Jia}, \bibinfo{person}{Shanshan Li}, \bibinfo{person}{Ying Wang}, \bibinfo{person}{Jun Ma}, \bibinfo{person}{Xiaoling Li}, \bibinfo{person}{Haoran Liu}, \bibinfo{person}{Ying Fu}, {and} \bibinfo{person}{Xiangke Liao}.} \bibinfo{year}{2024}\natexlab{}.
\newblock \showarticletitle{How to Pet a Two-Headed Snake? Solving Cross-Repository Compatibility Issues with Hera}. In \bibinfo{booktitle}{\emph{Proceedings of the 39th IEEE/ACM International Conference on Automated Software Engineering}}. \bibinfo{pages}{694--705}.
\newblock


\bibitem[Xu et~al\mbox{.}(2024)]%
        {xu2024cruxeval}
\bibfield{author}{\bibinfo{person}{Ruiyang Xu}, \bibinfo{person}{Jialun Cao}, \bibinfo{person}{Yaojie Lu}, \bibinfo{person}{Hongyu Lin}, \bibinfo{person}{Xianpei Han}, \bibinfo{person}{Ben He}, \bibinfo{person}{Shing-Chi Cheung}, {and} \bibinfo{person}{Le Sun}.} \bibinfo{year}{2024}\natexlab{}.
\newblock \showarticletitle{CRUXEval-X: A Benchmark for Multilingual Code Reasoning, Understanding and Execution}.
\newblock \bibinfo{journal}{\emph{arXiv preprint arXiv:2408.13001}} (\bibinfo{year}{2024}).
\newblock


\bibitem[Xue et~al\mbox{.}(2023)]%
        {xue2023acwrecommender}
\bibfield{author}{\bibinfo{person}{Zhipeng Xue}, \bibinfo{person}{Zhipeng Gao}, \bibinfo{person}{Xing Hu}, {and} \bibinfo{person}{Shanping Li}.} \bibinfo{year}{2023}\natexlab{}.
\newblock \showarticletitle{ACWRecommender: A Tool for Validating Actionable Warnings with Weak Supervision}. In \bibinfo{booktitle}{\emph{2023 38th IEEE/ACM International Conference on Automated Software Engineering (ASE)}}. IEEE, \bibinfo{pages}{1876--1880}.
\newblock


\bibitem[Xue et~al\mbox{.}(2024)]%
        {xue2024selfpico}
\bibfield{author}{\bibinfo{person}{Zhipeng Xue}, \bibinfo{person}{Zhipeng Gao}, \bibinfo{person}{Shaohua Wang}, \bibinfo{person}{Xing Hu}, \bibinfo{person}{Xin Xia}, {and} \bibinfo{person}{Shanping Li}.} \bibinfo{year}{2024}\natexlab{}.
\newblock \showarticletitle{SelfPiCo: Self-Guided Partial Code Execution with LLMs}. In \bibinfo{booktitle}{\emph{Proceedings of the 33rd ACM SIGSOFT International Symposium on Software Testing and Analysis}}. \bibinfo{pages}{1389--1401}.
\newblock


\bibitem[Yan et~al\mbox{.}(2024)]%
        {yan2024better}
\bibfield{author}{\bibinfo{person}{Jiwei Yan}, \bibinfo{person}{Jinhao Huang}, \bibinfo{person}{Chunrong Fang}, \bibinfo{person}{Jun Yan}, {and} \bibinfo{person}{Jian Zhang}.} \bibinfo{year}{2024}\natexlab{}.
\newblock \showarticletitle{Better Debugging: Combining Static Analysis and LLMs for Explainable Crashing Fault Localization}.
\newblock \bibinfo{journal}{\emph{arXiv preprint arXiv:2408.12070}} (\bibinfo{year}{2024}).
\newblock


\bibitem[Yang et~al\mbox{.}(2024b)]%
        {qwen2}
\bibfield{author}{\bibinfo{person}{An Yang}, \bibinfo{person}{Baosong Yang}, \bibinfo{person}{Binyuan Hui}, \bibinfo{person}{Bo Zheng}, \bibinfo{person}{Bowen Yu}, \bibinfo{person}{Chang Zhou}, \bibinfo{person}{Chengpeng Li}, \bibinfo{person}{Chengyuan Li}, \bibinfo{person}{Dayiheng Liu}, \bibinfo{person}{Fei Huang}, {et~al\mbox{.}}} \bibinfo{year}{2024}\natexlab{b}.
\newblock \showarticletitle{Qwen2 technical report}.
\newblock \bibinfo{journal}{\emph{arXiv preprint arXiv:2407.10671}} (\bibinfo{year}{2024}).
\newblock


\bibitem[Yang et~al\mbox{.}(2024a)]%
        {yang2024sweagent}
\bibfield{author}{\bibinfo{person}{John Yang}, \bibinfo{person}{Carlos~E Jimenez}, \bibinfo{person}{Alexander Wettig}, \bibinfo{person}{Kilian Lieret}, \bibinfo{person}{Shunyu Yao}, \bibinfo{person}{Karthik Narasimhan}, {and} \bibinfo{person}{Ofir Press}.} \bibinfo{year}{2024}\natexlab{a}.
\newblock \showarticletitle{Swe-agent: Agent-computer interfaces enable automated software engineering}.
\newblock \bibinfo{journal}{\emph{arXiv preprint arXiv:2405.15793}} (\bibinfo{year}{2024}).
\newblock


\bibitem[Yu et~al\mbox{.}(2023)]%
        {yu2023wavecoder}
\bibfield{author}{\bibinfo{person}{Zhaojian Yu}, \bibinfo{person}{Xin Zhang}, \bibinfo{person}{Ning Shang}, \bibinfo{person}{Yangyu Huang}, \bibinfo{person}{Can Xu}, \bibinfo{person}{Yishujie Zhao}, \bibinfo{person}{Wenxiang Hu}, {and} \bibinfo{person}{Qiufeng Yin}.} \bibinfo{year}{2023}\natexlab{}.
\newblock \showarticletitle{Wavecoder: Widespread and versatile enhanced instruction tuning with refined data generation}.
\newblock \bibinfo{journal}{\emph{arXiv preprint arXiv:2312.14187}} (\bibinfo{year}{2023}).
\newblock


\bibitem[Zelikman et~al\mbox{.}(2022)]%
        {zelikman2022star}
\bibfield{author}{\bibinfo{person}{Eric Zelikman}, \bibinfo{person}{Yuhuai Wu}, \bibinfo{person}{Jesse Mu}, {and} \bibinfo{person}{Noah Goodman}.} \bibinfo{year}{2022}\natexlab{}.
\newblock \showarticletitle{Star: Bootstrapping reasoning with reasoning}.
\newblock \bibinfo{journal}{\emph{Advances in Neural Information Processing Systems}}  \bibinfo{volume}{35} (\bibinfo{year}{2022}), \bibinfo{pages}{15476--15488}.
\newblock


\bibitem[Zhang et~al\mbox{.}(2024a)]%
        {zhang2024codeagent}
\bibfield{author}{\bibinfo{person}{Kechi Zhang}, \bibinfo{person}{Jia Li}, \bibinfo{person}{Ge Li}, \bibinfo{person}{Xianjie Shi}, {and} \bibinfo{person}{Zhi Jin}.} \bibinfo{year}{2024}\natexlab{a}.
\newblock \showarticletitle{CodeAgent: Enhancing Code Generation with Tool-Integrated Agent Systems for Real-World Repo-level Coding Challenges}.
\newblock \bibinfo{journal}{\emph{arXiv preprint arXiv:2401.07339}} (\bibinfo{year}{2024}).
\newblock


\bibitem[Zhang et~al\mbox{.}(2024d)]%
        {zhang2024diversity}
\bibfield{author}{\bibinfo{person}{Kexun Zhang}, \bibinfo{person}{Weiran Yao}, \bibinfo{person}{Zuxin Liu}, \bibinfo{person}{Yihao Feng}, \bibinfo{person}{Zhiwei Liu}, \bibinfo{person}{Rithesh Murthy}, \bibinfo{person}{Tian Lan}, \bibinfo{person}{Lei Li}, \bibinfo{person}{Renze Lou}, \bibinfo{person}{Jiacheng Xu}, {et~al\mbox{.}}} \bibinfo{year}{2024}\natexlab{d}.
\newblock \showarticletitle{Diversity empowers intelligence: Integrating expertise of software engineering agents}.
\newblock \bibinfo{journal}{\emph{arXiv preprint arXiv:2408.07060}} (\bibinfo{year}{2024}).
\newblock


\bibitem[Zhang et~al\mbox{.}(2023)]%
        {zhang2023lampr}
\bibfield{author}{\bibinfo{person}{Mengxiao Zhang}, \bibinfo{person}{Yongqiang Tian}, \bibinfo{person}{Zhenyang Xu}, \bibinfo{person}{Yiwen Dong}, \bibinfo{person}{Shin~Hwei Tan}, {and} \bibinfo{person}{Chengnian Sun}.} \bibinfo{year}{2023}\natexlab{}.
\newblock \showarticletitle{Lampr: Boosting the Effectiveness of Language-Generic Program Reduction via Large Language Models}.
\newblock \bibinfo{journal}{\emph{arXiv preprint arXiv:2312.13064}} (\bibinfo{year}{2023}).
\newblock


\bibitem[Zhang et~al\mbox{.}(2024c)]%
        {zhang2024lpr}
\bibfield{author}{\bibinfo{person}{Mengxiao Zhang}, \bibinfo{person}{Yongqiang Tian}, \bibinfo{person}{Zhenyang Xu}, \bibinfo{person}{Yiwen Dong}, \bibinfo{person}{Shin~Hwei Tan}, {and} \bibinfo{person}{Chengnian Sun}.} \bibinfo{year}{2024}\natexlab{c}.
\newblock \showarticletitle{LPR: Large Language Models-Aided Program Reduction}. In \bibinfo{booktitle}{\emph{Proceedings of the 33rd ACM SIGSOFT International Symposium on Software Testing and Analysis}}. \bibinfo{pages}{261--273}.
\newblock


\bibitem[Zhang et~al\mbox{.}(2024b)]%
        {autocoderover}
\bibfield{author}{\bibinfo{person}{Yuntong Zhang}, \bibinfo{person}{Haifeng Ruan}, \bibinfo{person}{Zhiyu Fan}, {and} \bibinfo{person}{Abhik Roychoudhury}.} \bibinfo{year}{2024}\natexlab{b}.
\newblock \showarticletitle{Autocoderover: Autonomous program improvement}. In \bibinfo{booktitle}{\emph{Proceedings of the 33rd ACM SIGSOFT International Symposium on Software Testing and Analysis}}. \bibinfo{pages}{1592--1604}.
\newblock


\bibitem[Zheng et~al\mbox{.}(2024)]%
        {zheng2024opencodeinterpreter}
\bibfield{author}{\bibinfo{person}{Tianyu Zheng}, \bibinfo{person}{Ge Zhang}, \bibinfo{person}{Tianhao Shen}, \bibinfo{person}{Xueling Liu}, \bibinfo{person}{Bill~Yuchen Lin}, \bibinfo{person}{Jie Fu}, \bibinfo{person}{Wenhu Chen}, {and} \bibinfo{person}{Xiang Yue}.} \bibinfo{year}{2024}\natexlab{}.
\newblock \showarticletitle{Opencodeinterpreter: Integrating code generation with execution and refinement}.
\newblock \bibinfo{journal}{\emph{arXiv preprint arXiv:2402.14658}} (\bibinfo{year}{2024}).
\newblock


\bibitem[Zhu and Zhou(2024)]%
        {zhu2024moss}
\bibfield{author}{\bibinfo{person}{Ming Zhu} {and} \bibinfo{person}{Yi Zhou}.} \bibinfo{year}{2024}\natexlab{}.
\newblock \showarticletitle{MOSS: Enabling Code-Driven Evolution and Context Management for AI Agents}.
\newblock \bibinfo{journal}{\emph{arXiv preprint arXiv:2409.16120}} (\bibinfo{year}{2024}).
\newblock


\bibitem[Zhu et~al\mbox{.}(2024a)]%
        {zhu2024domaineval}
\bibfield{author}{\bibinfo{person}{Qiming Zhu}, \bibinfo{person}{Jialun Cao}, \bibinfo{person}{Yaojie Lu}, \bibinfo{person}{Hongyu Lin}, \bibinfo{person}{Xianpei Han}, \bibinfo{person}{Le Sun}, {and} \bibinfo{person}{Shing-Chi Cheung}.} \bibinfo{year}{2024}\natexlab{a}.
\newblock \showarticletitle{DOMAINEVAL: An Auto-Constructed Benchmark for Multi-Domain Code Generation}.
\newblock \bibinfo{journal}{\emph{arXiv preprint arXiv:2408.13204}} (\bibinfo{year}{2024}).
\newblock


\bibitem[Zhu et~al\mbox{.}(2024b)]%
        {zhu2024deepseekcoder}
\bibfield{author}{\bibinfo{person}{Qihao Zhu}, \bibinfo{person}{Daya Guo}, \bibinfo{person}{Zhihong Shao}, \bibinfo{person}{Dejian Yang}, \bibinfo{person}{Peiyi Wang}, \bibinfo{person}{Runxin Xu}, \bibinfo{person}{Y Wu}, \bibinfo{person}{Yukun Li}, \bibinfo{person}{Huazuo Gao}, \bibinfo{person}{Shirong Ma}, {et~al\mbox{.}}} \bibinfo{year}{2024}\natexlab{b}.
\newblock \showarticletitle{DeepSeek-Coder-V2: Breaking the Barrier of Closed-Source Models in Code Intelligence}.
\newblock \bibinfo{journal}{\emph{arXiv preprint arXiv:2406.11931}} (\bibinfo{year}{2024}).
\newblock


\end{thebibliography}










\end{document}